%% file: driverFile.tex
\documentclass{tcibook}
\usepackage{fancyhea}
\usepackage{work}
\usepackage{bm}       
\usepackage{graphicx}
\usepackage{hyperref}      

\input workshopsymbols.tex      

\usepackage{chngcntr}
\counterwithout{figure}{chapter}

\begin{document}

\def\bibname{References}

\bibliographystyle{utphys}  

\raggedbottom

\pagenumbering{roman}

\parindent=0pt
\parskip=8pt
\setlength{\evensidemargin}{0pt}
\setlength{\oddsidemargin}{0pt}
\setlength{\marginparsep}{0.0in}
\setlength{\marginparwidth}{0.0in}
\marginparpush=0pt


\pagenumbering{arabic}

\renewcommand{\chapname}{chap:intro_}
\renewcommand{\chapterdir}{.}
\renewcommand{\arraystretch}{1.25}
\addtolength{\arraycolsep}{-3pt}


\input CF6/wgreportCF6.tex



\end{document}

%% file: workshopsymbols.tex
\newcommand{\nc}{\newcommand}  

\def\beq{\begin{equation}}
\def\eeq#1{\label{#1}\end{equation}}
\def\eeqn{\end{equation}}

\newenvironment{Eqnarray}%
   {\arraycolsep 0.14em\begin{eqnarray}}{\end{eqnarray}}
\def\beqa{\begin{Eqnarray}}
\def\eeqa#1{\label{#1}\end{Eqnarray}}
\def\eeqan{\end{Eqnarray}}

\nc{\ra}{\rightarrow}  
\nc{\slsh}{\slash\hspace*{-0.22cm}}
\def\Re{{\cal R \mskip-4mu \lower.1ex \hbox{\it e}\,}}
\def\Im{{\cal I \mskip-5mu \lower.1ex \hbox{\it m}\,}}

\nc{\vev}[1]{ \left\langle {#1} \right\rangle }
\nc{\bra}[1]{ \langle {#1} | }
\nc{\ket}[1]{ | {#1} \rangle }
\nc{\fb}{\,{\rm fb}^{-1}}
\nc{\ev}{{\rm eV}}
\nc{\kev}{{\rm keV}}
\nc{\Mev}{{\rm MeV}}
\nc{\gev}{{\rm GeV}}
\nc{\tev}{{\rm TeV}}
\nc{\mev}{{\rm MeV}}

\def\del{\partial}
\def\Dslash{\not{\hbox{\kern-4pt $D$}}}
\def\dslash{\not{\hbox{\kern-2pt $\del$}}}
\def\pslash{\not{\hbox{\kern-2pt $p$}}}
\def\ETmiss{ \not{\hbox{\kern-4pt $E$}}_T }

\def\msb{{\bar{\ssstyle M \kern -1pt S}}}



%% file: CF6/wgreportCF6.tex
 
\chapter*{CF6 Working Group Summary}

\renewcommand*\thesection{\arabic{section}}

{\center \bf \LARGE The Bright Side of the Cosmic Frontier:\\Cosmic Probes of
  Fundamental Physics\\}

\begin{center}\begin{boldmath}

\input CF6/authorlistCF6.tex

\end{boldmath}\end{center}


\input{CF6/executive-summary-v4.tex}
\section{CF6-A: Cosmic Rays, Gamma Rays, and Neutrinos}
\subsection{Cosmic rays and Particle Acceleration}
\input{CF6/cosmicrays.tex}
\input{CF6/accel.tex}

\input{CF6/neutrinos.tex}

\subsection{Cosmic Particles as Probes of Particle Physics Beyond Laboratory Energies}
\input{CF6/hadronicinteractions.tex}
\input{CF6/neutrinointeractions.tex}

\input{CF6/liv.tex}
\subsection{Rare and New Particles}
\input{CF6/antimatter.tex}
\input{CF6/pbh.tex}

\input{CF6/alp.tex}
\input{CF6/monopole.tex}
\input{CF6/qballs.tex}
\input{CF6/strangelet.tex}
\subsubsection{Quark anti-nuggets (Angela)}
\subsection{New Facilities}
\subsubsection{UHECR Experiments}
\input{CF6/auger-upgrade.tex}
\input{CF6/telescope-array.tex}

\input{CF6/JEM-EUSO.tex}

\input{CF6/CR-radio.tex}
\input{CF6/CR-radar.tex}

\input{CF6/gammaexp}
\subsubsection{Neutrino Experiments}
\input{CF6/icecube.tex}

\input{CF6/pingu.tex}
\input{CF6/nu-askaryan.tex}

\section{CF6-B: The Matter Of the Cosmological Asymmetry}
\input{CF6/matter-assymmetry.tex}

\section{CF6-C: Exploring the Basic Nature of Space and Time}
\input{CF6/holometer.tex}
\input{CF6/TQ.tex}



\bibliography{CF6/CF6}{}


%% file: CF6/authorlistCF6.tex


\begin{center}

\begin{large} {\bf Conveners: J.J.~Beatty, A.E.~Nelson, A.~Olinto, G.~Sinnis} \end{large}

A.~U.~Abeysekara, 
L.A.~Anchordoqui, 
T.~Aramaki, 
J.~Belz, 
J.H.~Buckley, 
K.~Byrum, 
R.~Cameron, 
M-C.~Chen, 
K.~Clark,
A.~Connolly, 
D.F.~Cowen, 
T.~DeYoung, 
P.~von~Doetinchem 
J.~Dumm, 
M.~Errando, 
G.~Farrar, 
F.~Ferrer, 
L.~Fortson, 
S.~Funk, 
D.~Grant, 
S.~Griffiths, 
A.~Gro\ss,
C.~Hailey, 
C.~Hogan, 
J.~Holder, 
B.~Humensky, 
P.~Kaaret, 
S.R.~Klein, 
H.~Krawczynski, 
F.~Krennrich, 
K.~Krings,
J.~Krizmanic, 
A.~Kusenko, 
J.~T.~Linnemann, 
J.~H.~MacGibbon, 
J.~Matthews, 
A.~McCann, 
J.~Mitchell, 
R.~Mukherjee, 
D.~Nitz, 
R.A.~Ong, 
M.~Orr, 
N.~Otte,
T.~Paul, 
E.~Resconi,
M.~A.~Sanchez-Conde, 
P.~Sokolsky, 
F.~Stecker, 
D.~Stump, 
I.~Taboada, 
G.B.~Thomson, 
K.~Tollefson, 
P.~von~Doetinchem, 
T.~Ukwatta, 
J.~Vandenbroucke, 
V. Vasileiou, 
V.V.~Vassileiv, 
T.J.~Weiler, 
D.A.~Williams, 
A.~Weinstein, 
M.~Wood, 
B. Zitzer 

\end{center}


%% file: CF6/executive-summary-v4.tex





\section{Executive Summary} 

Over the past decade we have witnessed a revolution in our
understanding of the high-energy universe.  Some of the key
discoveries have been:
\begin{itemize}
\item Supernovae have been shown to be a source of Galactic cosmic
  rays \cite{ackermann2013a,Giuliani:2011nx,acciarri2011}.
\item Very high energy neutrinos that are likely to be astrophysical
  in origin have been observed \cite{Aartsen:2013bka}.
\item The GZK suppression \cite{Greisen:1966jv,Zatsepin:1966jv} in the cosmic-ray flux above $10^{19.5}$ eV
  has been observed \cite{Abbasi:2007sv,Abraham:2008ru,AbuZayyad:2012ru,Abraham:2010mj}.
\item The positron fraction of the cosmic rays has been measured up to
  300 GeV and provides solid evidence for a high-energy primary source
  of positrons in the Galaxy, either from dark matter annihilation or
  astrophysical processes \cite{adriani2009,acciarri2011,Aguilar:2013qda,yuksel2009}.
\item Many sites of astrophysical particle acceleration have been
  directly observed, from supermassive black holes and merging neutron
  stars, to rapidly spinning neutron stars and supernova remnants in
  our Galaxy \cite{Aharonian:2008zza,funk2012,rieger2013,tevcat}.
\end{itemize}

These discoveries have been driven by the current generation of
experiments: the IceCube neutrino detector at the South Pole \cite{Collaboration:2011ym,Halzen:2010yj}, the
Fermi gamma-ray observatory \cite{Atwood2009} and the PAMELA \cite{menn2013} and AMS experiments \cite{kounine2012}
orbiting the earth, the High Resolution FlyÕs Eye \cite{boyer2002} and Pierre Auger \cite{Abraham:2010mj}
ultra-high-energy cosmic ray experiments, and the H.E.S.S. \cite{aharonian2005}, VERITAS \cite{weekes2002},
MAGIC \cite{aleksic2011} and Milagro \cite{atkins2003} experiments in very high energy gamma rays.  Looking
forward, a new generation of instruments with greater sensitivity and
higher resolution hold the promise of making large advances in our
understanding of astrophysical processes and the fundamental physics
studied with astrophysical accelerators.  The goals for the coming
decade are:
\begin{itemize}
\item Determine the origin of the highest energy particles in the
  universe and understand the acceleration processes at work
  throughout the Universe.
\item Measure particle cross sections at energies unattainable in
  Earth-bound accelerators.
\item Measure the highest energy neutrinos that arise from
  interactions of the ultra-high-energy cosmic rays with the microwave
  background radiation.
\item Measure the extragalactic background light between 1 and 100
  microns to understand the star formation history of the Universe.
\item Measure the mass hierarchy of the neutrinos.
\item Search for physics beyond the standard model encoded in cosmic
  messengers as they cross the Universe.
\item Understand the origins of the matter antimatter asymmetry of the
  Universe.
\item Probe the fundamental nature of spacetime.
\end{itemize}

In many of these areas future progress will depend upon either the
detailed understanding of particle acceleration in the universe or the
development of methods for controlling systematic errors introduced by
our lack of understanding of these processes.  High-resolution gamma
ray measurements (spectral, angular, and temporal) of many objects and
classes of objects are needed to find the source invariant physics
that is the signal for physics beyond the standard model.  Such
measurements in conjunction with measurements at other wavelengths and
with measurements of cosmic rays, neutrinos, and gravity waves will
enable us to understand Nature's particle accelerators.

\underline{\bf Recommendations:}
\begin{itemize}
\item Significant U.S. participation in the CTA project \cite{cta2011}.
  U.S. scientists developed the imaging atmospheric Cherenkov
  technique.  Continued leadership in this area is possible with the
  development of novel telescope designs.   A U.S. proposal to more than double
  the number of mid-scale telescopes would result in a sensitivity gain of
  2-3 significantly improving the prospects for the
  indirect detection of dark matter, understanding particle
  acceleration processes, and searching for other signatures of
  physics beyond the standard model.
\item Simultaneous operation of Fermi, HAWC, and VERITAS.
  Understanding particle acceleration and separating astrophysical
  processes from physics beyond the standard model requires
  observations over a broad energy range.  The above three instruments
  will provide simultaneous coverage from 30 MeV to 100 TeV.  HAWC and
  VERITAS will simultaneously view the same sky enabling prompt follow
  up observations of transient phenomena.
\item Construction of the PINGU neutrino detector \cite{aartsen2013} at the South Pole.
  U.S. scientists have been leaders in the field of high-energy
  neutrino observations.  PINGU, by densely instrumenting a portion of
  the IceCube Deep Core array, will lower the energy threshold for
  neutrinos to a few GeV.  This will allow for a measurement of the
  neutrino mass hierarchy using atmospheric neutrinos.
\item Continued operation of the Auger and TA air shower arrays with
  upgrades to enhance the determination of the composition and
  interactions of cosmic rays near the energy of the GZK suppression,
  and flight of the JEM-EUSO mission \cite{Adams:2013vea,TheJEM-EUSO:2013vea} to extend observations of the
  cosmic ray spectrum and anisotropy well beyond the GZK region.
\item Construction of a next-generation ultra-high energy GZK neutrino
  detector either to detect GZK neutrinos \cite{berezinsky2006} and constrain the
  neutrino-nucleon cross section at these energies \cite{Klein2013,hooper2002,Borriello2008,Connolly2011,Romero2009}, or rule out all
  but the most unfavorable parts of the allowed parameter space.
\end{itemize}

\subsection{Ultra-High-Energy Cosmic Rays}

HiRes  \cite{Abbasi:2007sv}, Auger  \cite{Abraham:2008ru,Abraham:2010mj}, and the Telescope Array (TA) \cite{AbuZayyad:2012ru} have established the
existence of a suppression of the spectrum at the highest energies
(above $\sim6\times 10^{19}$ eV ) as predicted by Greisen, Zatsepin,
and Kuzmin (GZK) in 1966 \cite{Greisen:1966jv,Zatsepin:1966jv}.  The GZK suppression is an example of the
profound links between different regimes of physics, connecting the
behavior of the highest-energy particles in the Universe to the cosmic
microwave background radiation, and can be explained by the sub-GeV
scale physics of photo-pion production occurring in the extremely
boosted relativistic frame of the cosmic ray.  A similar phenomenon
occurs for primary nuclei due to excitation of the giant dipole
resonance, resulting in photo-disintegration.  For iron nuclei, this
occurs at about the same energy per particle as the photo-pion process
does for protons.

The composition of cosmic rays and their interactions with air nuclei
may be probed by studies of the depth of shower maximum, $X_{max}$ \cite{ellsworth1982, baltrusaitis1984,Collaboration:2012wt}.
The mean value of $X_{max}$ rises linearly as a function of the log of
the energy, and depends on the nature of the primary particle, the
depth of its first interaction, and the multiplicity and inelasticity
of the interactions as the shower evolves.  Lower energy observations
of $X_{max}$ indicate that the composition becomes lighter as the
energy increases toward $\sim10^{18.3}$ eV \cite{FlysEye93,gaisser1993,bird1994}, which suggests that
extragalactic cosmic rays are mainly protons.  However, at higher
energies the Auger ObservatoryÕs high quality, high-statistics sample
exhibits the opposite trend, along with a decreasing spread in
$X_{max}$ with increasing energy \cite{Abraham:2010mj,Abreu:2013env}.  Using current simulations and
hadronic models tuned with LHC forward data, this implies the
composition is becoming gradually heavier above $10^{18.5}$ eV.  A
trend toward heavier composition could reflect the apparent GZK
suppression being in fact the endpoint of cosmic acceleration in which
there is a maximum magnetic rigidity for acceleration, resulting in
heavy nuclei having the highest energy per particle.

Cosmic rays can be used to probe particle physics at energies far
exceeding those available at the LHC.  An alternative explanation for
the observed behavior of $X_{max}$ is a change in particle
interactions not captured in event generators tuned to LHC data.
Auger measurements using three independent methods find that these
models do not describe observed showers well.  For example, the
observed muon content of showers measured in hybrid events at Auger is
a factor 1.3 to 1.6 higher larger than predicted \cite{AugerMuonICRC13}.  TA also observes a
calorimetric energy that is about 1.3 times higher than that inferred
from their surface detector using these models.  An example of a novel
phenomenon that may explain these observations is the restoration of
chiral symmetry in QCD  \cite{Farrar:2013sfa}.

A critical step in fully understanding the $X_{max}$ observations is
to identify and correct the deficiencies in the beyond-LHC physics
used in modeling showers.  This requires continued operation of
current hybrid detectors such as Auger and TA with upgrades to enable
improved multiparameter studies of composition and interactions on a
shower-by-shower basis.  Enhancements of the surface detectors are
particularly valuable because of the tenfold higher duty cycle than
for fluorescence or hybrid operation.

Observations from space can extend studies of the spectrum and
anisotropy beyond the GZK region with high statistics.  Current
ground-based observatories have observed hints of correlation of
cosmic ray arrival directions with the local distribution of matter \cite{Cronin:2007zz,Abraham:2007si,Abreu:2010ab,AugerICRC-HL12,AbuZayyad:2012hv,TinyakovTAanisoICRC13}, but higher statistics trans-GZK observations are required to identify
sources. In addition, the question of whether the spectrum flattens
again above the GZK suppression or continues to fall will distinguish
between the GZK and acceleration limit scenarios.  The JEM-EUSO
mission has an instantaneous collecting area of $\sim$40  \cite{Adams:2013vea,TheJEM-EUSO:2013vea} times that of
existing ground-based detectors and, taking duty cycle into account,
will increase the collecting area above the GZK suppression energy by
nearly an order of magnitude.

\subsection{Neutrinos}

IceCube has recently reported the detection
of two neutrinos with energies above 1 PeV and 26 events above 30 TeV
with characteristics that point to an astrophysical origin \cite{Aartsen:2013bka}.  These
exciting results herald the beginning of the era of high-energy
neutrino astronomy, and initiate the study of ultra-long baseline
high-energy neutrino oscillations.  Neutrino data complements
observations of cosmic rays and gamma rays due to their origin in the
decays following high-energy hadronic interactions and their weak
couplings.  Several acute issues in particle physics and astrophysics
can be addressed by neutrino experiments.

\underline{\em GZK Neutrinos} Neutrinos are produced by the weak
decays of the mesons and neutrons produced in the interaction of UHE
cosmic rays with the CMB \cite{berezinsky2006}.  The production of these neutrinos takes
place via well-known physics at high Lorentz boost, so robust
predictions of the neutrino flux can be made.  This flux depends on
the composition of the primary cosmic rays and the evolution of the
cosmic ray source density with redshift.  Unlike many searches, there
is a lower limit on the expected flux.  Current detectors such as
IceCube \cite{abbasi2011,ishihara2012}, Auger \cite{Abreu:2011zze}, RICE \cite{Kravchenko:2011im} and ANITA \cite{Hoover:2010qt,barwick2006} 
have begun to probe the highest predicted values of the neutrino flux.  Next generation experiments
such as ARA, ARIANNA, and EVA can increase our sensitivity by about
two orders of magnitude, and will either detect GZK neutrinos or rule
out much of the parameter space.  If GZK neutrinos are detected, the
event rate as a function of zenith angle can be used to measure the
neutrino-nucleon cross section and constrain models with enhanced
neutrino interactions at high energy.

\underline{\em Atmospheric Neutrinos and the Neutrino Mass Hierarchy}
PINGU \cite{aartsen2013} is a proposed high-density infill of the IceCube detector with a
reduced energy threshold of a few GeV, employing the rest of IceCube
as an active veto.  PINGU has sensitivity to atmospheric $\nu_{\mu}$
over a range of values of L/E spanned by the variation in the distance
to the production region as a function of zenith angle and the energy
spectrum of atmospheric neutrinos.  Preliminary studies of PINGU
indicate that over these values of L/E, atmospheric $\nu_{\mu}$
oscillations can be used to determine the neutrino mass hierarchy with
3-5$\sigma$ significance with two years of data with a 40 string
detector \cite{aartsen2013} .

\underline{\em Supernova Neutrinos} The Long Baseline Neutrino
Experiment \cite{Adams:2013qkq} will consist of a large (10-35 kTon) liquid argon time
projection chamber located at 4850 feet depth at the Homestake mine in
South Dakota. Collective oscillations of neutrinos as they traverse
the neutrinosphere lead to a spectral swap that leaves a signature in
the energy spectrum of electron neutrinos that is dependent upon the
neutrino mass hierarchy \cite{duan2006,duan2007,duan2010,dasgupta2009}.  In the event of a Galactic supernova,
sufficient numbers of events would be detected to both measure the
neutrino mass hierarchy and to elucidate the supernova mechanism.

\subsection{Gamma Rays}

High-energy gamma rays provide a unique view into the most extreme
environments in the universe, allowing one to probe particle
acceleration processes and the origin of the Galactic and
extragalactic cosmic rays.  Active galactic nuclei (AGN), supermassive
black holes emitting jets of highly relativistic particles along their
rotation axis, have been shown to be sites of particle acceleration \cite{Aharonian:2008zza,funk2012,rieger2013,tevcat}.
Outstanding issues in the acceleration processes include: the nature
of the accelerated particles (hadronic or leptonic), the role of shock
acceleration versus magnetic reconnection, and the formation and
collimation of astrophysical jets \cite{boettcher2007,Bykov:2012ca}.  Answers to these questions will
come from higher resolution measurements in the GeV-TeV regime,
multi-wavelength campaigns with radio, x-ray, and gamma-ray
instruments, multi-messenger observations with gamma rays, ultra-high
energy cosmic rays, neutrinos, and potentially gravitational
waves. Understanding these extreme environments and how they
accelerate particles is of fundamental interest.  In addition, these
high-energy particle beams, visible from cosmologically interesting
distances, allow us to probe fundamental physics at scales and in ways
that are not possible in earth-bound particle accelerators.

Recently, Fermi \cite{ackermann2013a} and AGILE \cite{Giuliani:2011nx} measured the energy spectra of the supernova
remnants W44 and IC44.  The decrease in the gamma-ray flux below the
pion mass in these sources is clear evidence for hadronic
acceleration.  This is the clearest evidence to date that some
Galactic cosmic rays are accelerated in supernova remnants.  The
detection of high-energy neutrinos from a cosmic accelerator would be
a smoking gun signature of hadronic acceleration.  In the absence of
multi-messenger signals multi-wavelength energy spectra (x-ray to TeV)
can test both leptonic models, where the high-energy emission is
derived from inverse Compton scattering of the x-ray synchrotron
emission, and hadronic models, where gamma rays result from pion
decays or proton synchrotron radiation.

\underline{\em Backgrounds to dark matter searches.} Understanding the
origin of the Galactic VHE gamma rays is critical in the
interpretation of some signatures of dark matter annihilation.  The
recent results from PAMELA \cite{adriani2009} and AMS \cite{Aguilar:2013qda} of the increasing positron fraction
with energy is a clear signal that the current model of secondary
production and transport through the Galaxy is not complete.  There are
three potential explanations for this signal: a new astrophysical
source of positrons \cite{Hooper:2008kg,Linden:2013mqa}, modified propagation of cosmic rays or secondary
production in the source \cite{Zirakashvili:2011sc,PhysRevD.82.023009}, or dark matter annihilation \cite{turner1990}.  An
astrophysical source of positrons, pulsar wind nebula (PWN), is now
known to also lead to an increasing positron fraction at high
energies.  Observations of the Geminga PWN in the TeV band by Milagro \cite{abdo2009b}
have been used to normalize the flux of positrons in our local
neighborhood from this source.  The calculated positron fraction is an
excellent match to the data \cite{yuksel2009}.  Similarly, in searching for dark matter
signatures from the Galactic center or galaxy clusters one must
understand and measure the more standard astrophysical processes that
may lead to signatures that are similar in nature to those expected
from dark matter annihilation.

\underline{\em Extragalactic background light.} In addition to
advancing our understanding of particle acceleration and astrophysical
backgrounds for dark matter searches, the intense gamma-ray beams
generated by AGN and gamma-ray bursts can be used to probe the
intervening space and search for physics beyond the standard
model. Some areas that can be studied include: measuring the
extragalactic background light (EBL), using the EBL to measure the
intergalactic magnetic fields, and searching for axion-like particles
(ALPs) (note, ALPs are not the QCD axion).  The EBL pervades the
universe and is the sum of all the light generated by stars and the
re-radiation of this light in the infrared band by dust \cite{Stecker:2012ta,dominguez2011,dwek2013}
and is therefore sensitive to the star formation history of the universe.
The production of electron-positron pairs from photon-photon
scattering of the EBL with high-energy gamma rays leads to an energy
dependent attenuation length for high-energy gamma rays \cite{Gould:1967zza}.  At the same
time, cosmic rays (if they are accelerated by AGN) will interact with
EBL and CMB along the line of sight and generate secondary gamma rays
at relatively close distances of the observer \cite{Aharonian:1993vz}.  This absorption, with
the inclusion of secondary gamma rays, can be used to measure or
constrain the EBL \cite{dominguez2011,dwek2013}.  Lower limits on the EBL can be established from
galaxy counts \cite{Stecker:2012ta}.  A signature of new physics would be an inconsistency
between these lower limits and the measurements of the TeV spectra
from AGN.  Such a discrepancy could be explained either by the
secondary production of gamma rays from cosmic rays produced at AGN or
by photon-alp mixing mediated by the intergalactic magnetic fields.

\underline{\em Intergalactic magnetic fields.} The origin of the
Galactic magnetic fields remains a mystery.  While astrophysical
dynamos can efficiently amplify pre-existing magnetic fields, the
generation of a magnetic field is difficult \cite{Kandus:2010nw}. The strength of the
magnetic fields in the voids between galaxy clusters should be similar
to the primordial magnetic field. Measurement of AGN spectra and time
delays in the GeV-TeV region have been used to set both lower and
upper bounds on the strength of the IGMF \cite{Plaga:1994hq,Ando:2010rb}.  Current bounds are model
dependent and span a large range of values for the magnetic field \cite{Essey:2011wv}.
Improved measurements of the EBL, the measurement of variability from
distant AGN, and improved determinations of the energy spectra (and
understanding of the intrinsic AGN spectra) are needed to
significantly improve these limits.

\underline{\em Tests of Lorentz invariance violation.}  Experimentally
probing Planck-scale physics, where quantum gravitational effects
become large, is challenging.  A unique signature of such affects
would be the violation of Lorentz invariance (a natural though not
necessary property of theories of quantum gravity) \cite{saslow1998,rovelli2008,solodukhin2011}.  
Short, intense pulses of gamma rays from distant objects such as gamma ray bursts and
active galaxies, provide a laboratory to search for small, energy
dependent, differences in the speed of light.  Current limits have
reached the Planck scale if the energy dependence of the violation is
linear \cite{abdo2009} and $6.4\times10^{10}$ GeV if the violation is quadratic \cite{abromowski2011} 
with energy.  Future instruments could improve upon these limits by at
least a factor of 10 and 50 (respectively) and significantly increase
the sample size used to search for these affects.

\underline{\em TeV Gamma-Ray Instruments.} Ground-based TeV
instruments fall into two classes: extensive air shower (EAS) arrays,
capable of simultaneously viewing the entire overhead sky, and imaging
atmospheric Cherenkov telescopes (IACTs), pointed instruments with
high sensitivity and resolution.  Current IACTs are VERITAS, MAGIC,
and HESS, while the HAWC (under construction) and Tibet arrays are the
only operating EAS arrays.  The next generation of IACT, known as CTA,
will consist of a large array of IACTs with roughly an order of
magnitude greater sensitivity than current instruments.  It is
expected to discover over 1000 sources in the TeV band \cite{cta2011, cta2013}.  
The U.S. portion of the collaboration is proposing to more than double the number of mid-sized telescopes 
to over 50 using a novel optical design that will improve the
sensitivity of CTA by a factor of 2-3 (a result of improved angular resolution of the new telescopes 
and an increase in the number of telescopes).

\subsection{Baryogenesis}

According to standard cosmology, the current preponderance of matter
arises during the very early universe from an asymmetry of about one
part per hundred million between the densities of quarks and
antiquarks. This asymmetry must have been created after inflation due
to some physical process known as baryogenesis.  Baryogenesis requires
extending the standard model of particle physics, via some new physics
which must couple to standard model particles and which must be
important during or after the end of inflation. Constraints on the
inflation scale and on the reheat temperature at the end of inflation
give an upper bound on the relevant energy scale for baryogenesis. The
new physics must violate CP (symmetry between matter and antimatter)
as the CP violation in the standard model is insufficient \cite{Gavela:1994dt}. There are a
very large number of theoretical baryogenesis proposals. Some
well-motivated possibilities are:

\underline{\em Leptogenesis} Theoretically, baryon number violation at
high temperatures is rapid in the standard model via nonperturbative
electroweak processes known as sphalerons. Sphalerons conserve the
difference between baryon and lepton numbers, leading to the idea of
leptogenesis \cite{Fukugita:1986hr}. The decay of very heavy neutrinos in the early universe
could occur in a CP violating way, creating a lepton asymmetry that
the sphalerons convert into a baryon asymmetry. Whether and how
leptogenesis could occur can be clarified by intensity frontier
experiments (CP violation and lepton number violation in neutrino
physics) \cite{DiBari:2012fz}, cosmic frontier experiments (most leptogenesis proposals
require a high inflation scale which can be constrained in cosmic
microwave background measurements), and energy frontier experiments \cite{Pilaftsis:2003gt}
(because whether or not other new physics exists has important
implications for leptogenesis).

\underline{\em Affleck-Dine baryogenesis} In supersymmetric theories
condensates of the scalar partners of quarks and leptons have
relatively low energy density and are likely to be present at the end
of inflation \cite{Affleck:1984fy}. The subsequent evolution and decay of the condensates
can produce the baryon asymmetry, and in some models, the dark
matter \cite{Kusenko:1997si,Kusenko:1997vp}. The dark matter could be macroscopic lumps of scalar quarks or
leptons known as Q-balls, which have unusual detection signatures in
cosmic frontier experiments \cite{Kusenko:2005du}. A critical test of this theory is the
search for supersymmetry, primarily via collider experiments.
Instruments such as HAWC and IceCube will be sensitive to extremely
low fluxes of Q-balls - over two orders of magnitude lower than
current limits.  While the non-observation of Q-balls can not rule out
the Affleck-Dine mechanism, the observation of a Q-ball would have a
profound impact on our understanding of the universe.

\underline{\em Electroweak baryogenesis} New non-standard model
particles that are coupled to the Higgs boson can provide a first
order phase transition for electroweak symmetry breaking, which
proceeds via nucleation and expansion of bubbles of broken phase. CP
violating interactions of particles with the expanding bubble walls
can lead to a CP violating particle density in the symmetric phase,
which can be converted by sphalerons into the baryon asymmetry \cite{Cohen:1992zx,Turok:1990zg}. The
baryons then enter the bubbles of broken phase where the sphaleron
rate is negligible. This scenario can be definitively tested via the
search for new particles in high-energy colliders, and via
nonvanishing electric dipole moments for the neutron and for
atoms. Non-standard CP violation in $D$ and $B$ physics may appear in
some models. A non standard Higgs self-coupling is a generic
consequence. The first order phase transition in the early universe
can show up via relic gravitational waves.

Other experiments that can shed light on baryogenesis include searches
for baryon number violation. Neutron anti-neutron oscillations violate
the difference between baryon and lepton numbers, and most
leptogenesis scenarios will not work if such processes are too
rapid. Proton decay would provide evidence for Grand Unification,
which would imply the existence of heavy particles whose decay could
be responsible for baryogenesis, and which is a feature of some
leptogenesis and supersymmetric models.

\subsection{Fundamental nature of spacetime}

Quantum effects of space-time are predicted to originate at the Planck
scale. In standard quantum field theory, their effects are strongly
suppressed at experimentally accessible energies, so space-time is
predicted to behave almost classically, for practical purposes, in
particle experiments. However, new quantum effects of geometry
originating at the Planck scale from geometrical degrees of freedom
not included in standard field theory may have effects on macroscopic
scales \cite{Jacobson:1995ab,Verlinde:2010hp} that could be measured by laser interferometers such as the
holometer \cite{Hogan:2012ib}. In addition, cosmic particles of high energies can probe
departures from Lorenz invariance \cite{amelino1998} and the existence of
extra-dimensions on scales above the LHC.

\subsection{Overview of the Report}

The Cosmic Frontier number 6, named Cosmic Probes of Fundamental
Physics, was charged with summarizing current knowledge and
identifying future opportunities (both experimental and theoretical)
in the use of astrophysical probes of fundamental physics. Because of
the breadth of this area of research CF6 has been subdivided into 3
main topical areas: 

\begin{itemize}
\item	CF6-A Cosmic Rays, Gamma Rays and Neutrinos (conveners: Gus Sinnis, Tom Weiler)
\item	CF6-B The Matter of the Cosmological Asymmetry (convener: Ann Nelson)
\item	CF6-C Exploring the Basic Nature of Space and Time (conveners: Aaron Chou, Craig Hogan)
\end{itemize}

We received 14 whitepapers addressing the current challenges and
future opportunities in each of these fields as they relate to the
other frontiers of high energy physics.

\subsubsection{CF6-A Cosmic Rays, Gamma Rays and
  Neutrinos}

Cosmic rays, gamma rays, and neutrinos from astrophysical sources can
be used to probe fundamental symmetries, particle interactions at
energies not attainable on Earth, and understand the nature of
particle acceleration in the cosmos. CF6-A encompasses the following
topical areas: The Physics of Interactions Beyond Laboratory Energies
(hadronic and weak interactions at high energies) Cosmic Particle
Acceleration (origin of the cosmic radiation from GeV to ZeV,
astrophysical backgrounds to fundamental physics) Cosmic Standard
Model (SM) Particles as Probes of Fundamental Physics (violation of
Lorentz invariance, extra dimensions, axions) New Particles
(anti-nuclei, strangelets, Q-Balls, primordial black holes) Neutrino
Physics from Astrophysics (in conjunction with the Intensity Frontier
IF3: neutrino mass hierarchy, leptonic CP violation, neutrino
interactions in supernovae, non-standard neutrino interactions with
matter)

 This field is ripe for new discoveries. Recent examples include the
 detection by the Fermi and AGILE space missions of a pion-zero decay
 produced gamma-ray signal in supernova remnants W44 and IC44
 \cite{FermiW44}; the first detection of 10 TeV to PeV neutrinos by
 Icecube \cite{BertErnie}; and the precise positron fraction
 measurement of the Alpha Magnetic Spectrometer
 \cite{AMSpositrons}. 

\subsubsection{CF6-B The
  Matter of the Cosmological Asymmetry}

The cosmological asymmetry between baryons and anti-baryons is one of
the strongest pieces of evidence for physics beyond the standard
model. There are diverse theoretical proposals for new physics at a
range of energy scales, most of which include new sources of CP
violation, and either additional baryon or lepton number
violation. There are plausible theories with implications for
experiments at all three frontiers. Leptogenesis theories predict new
CP violation in the neutrino sector. Grand Unified theories predict
proton decay. Electroweak baryogenesis models predict a cosmological
phase transition which leaves traces in gravitational wave and CMB
experiments, and new collider physics at or below a TeV.

We survey what we may expect to learn about the origin of matter from
searches at ongoing, planned and proposed facilities such as the LHC,
ILC, CLIC, muon collider, B-factories, neutrino experiments,
gravitational wave experiments and CMB experiments.

\subsubsection{CF6-C
  Exploring the Basic Nature of Space and Time}

Space and time may
be quantized at the Planck scale. Effects from this extremely
high-energy realm may be studied through gravitational waves from the
early universe and holographic noise.

We survey what we may expect to learn about the GUT and Planck energy
scale with gravitational waves and holographic noise. Survey the
theory and ongoing, planned, and proposed gravitational wave and
holographic noise experiments. Examples of relevant experiments
include the Holometer, LIGO, LISA, and CMB probes of early universe
gravitational wave (covered in Dark Energy and CMB.)



%% file: CF6/cosmicrays.tex
\subsubsection{Ultra high energy cosmic ray origins}

Ultra-high energy cosmic rays (UHECRs), now commonly taken to be CRs
with energies $> 6 \times 10^{19}$~eV, were first reported just over
50 years ago by John Linsley~\cite{Linsley:1963km}.  These are the
only particles with energies exceeding those available at terrestrial
accelerators. The Large Hadron Collider (LHC) will reach an equivalent
fixed-target energy of $10^{17}$ eV, whereas UHECRs have been observed
with energies in excess of $10^{20}$~eV. With UHECRs one can conduct
particle physics measurements up to two orders of magnitude higher in
the lab frame, or one order of magnitude higher in the center-of-mass
frame, than the LHC energy reach. As discussed in more detail below,
the properties of UHECR air showers appear to be inconsistent with
models which are tuned to accelerator measurements; one possible
explanation is that new physics intervenes at energies beyond the LHC
reach. UHECR experiments are the only way to access this energy range
and make detailed measurements of air showers in order to address this
question.  It is worth noting that cosmic ray experiments have already
yielded particle physics results at energies far exceeding those
accessible to the LHC, one of the latest being a measurement of the
$p$-air cross-section at $\sqrt{s} = 57~{\rm
TeV}$~\cite{Collaboration:2012wt}, a result which excludes some
hadronic models' extrapolations beyond LHC energies.

The two largest currently operating UHECR observatories are the Pierre
Auger Observatory in the Southern hemisphere, covering an area of
3000~km$^2$, and the Telescope Array (TA) in the Northern hemisphere,
covering about 700~km${^2}$.  Both observatories employ hybrid
detection techniques, sampling cosmic ray air shower particles as they
arrive at the Earth's surface and also detecting the fluorescence
light produced when UHECR air showers excite atmospheric nitrogen, for
the $\sim 10$\% of events arriving on dark, moonless nights.  Both
Auger and TA feature ``low energy'' extensions, which will provide an
overlap with the LHC energy regime, while also allowing measurements
in the galactic-to-extragalactic transition region.

The most important result so far from the present generation of
observatories is the conclusive evidence that the UHECR flux drops
precipitously at high energy. 
The
discovery of a suppression at the end of the cosmic ray spectrum was
first reported by HiRes and Auger~\cite{Abbasi:2007sv,Abraham:2008ru}
and later confirmed by TA~\cite{AbuZayyad:2012ru}; by now the
significance is well in excess of 20$\sigma$ compared to a continuous
power law extrapolation beyond the ankle
feature~\cite{Abraham:2010mj}.  This suppression is consistent with
the GZK prediction that interactions with cosmic background photons
will rapidly degrade UHECR energies~\cite{Greisen:1966jv}.
Intriguingly, however, there are also indications that the source of
the suppression may be more complex than originally anticipated.


Lower energy observations of the elongation rate (the rate of change with energy of the mean depth-of-shower-maximum, $X_{max}$)~\cite{FlysEye93,HiResMIA01,Tunka11,Yakutsk11}, indicate that the composition becomes lighter as energy increases toward $\sim 10^{18.3}$~eV, fueling a widespread supposition that extragalactic cosmic rays are primarily protons. At the highest energies however the situation remains ambiguous. HiRes and TA observe an elongation rate consistent with a light, unchanging composition, supporting the model in which the highest energy cosmic rays are protons and the energy spectrum is shaped by interactions with the CMBR~\cite{Allard:2011aa,AugerMuonICRC13}. However the Auger Observatory's data exhibits a decreasing elongation rate as well as a decreasing spread in $X_{max}$ with increasing energy. Interpreted with present shower simulations, this implies that the composition is becoming gradually heavier beginning around $5 \times 10^{18}$~eV~\cite{Abraham:2010yv,Abreu:2013env}. If true, this would have important implications for the astrophysics of the sources. A trend toward heavier composition could reflect the endpoint of cosmic acceleration, with heavier nuclei dominating the composition near the end of the spectrum --- which coincidentally falls off near the expected GZK cutoff region~\cite{Aloisio:2009sj}. In this scenario, the suppression would constitute an imprint of the accelerator characteristics rather than energy loss in transit. It is also possible that a mixed or heavy composition is emitted from the sources, and photodisintegration of nuclei and other GZK energy losses suppress the flux~\cite{Allard:2011aa}.


An alternative possibility for the origin of the break in the
elongation rate could be even more interesting: this feature might
arise from some change in the particle interactions at UHE not
captured by event generators tuned to LHC and other accelerator data.
Adding weight to this possibility are the Auger measurements, using
three independent methods, showing that existing hadronic interaction
models do not simultaneously fit all shower observables.  For example,
the actual hadronic muon content of UHE air showers measured in hybrid
events is a factor $1.3-1.6$ larger than predicted by models tuned to
LHC data~\cite{AugerMuonICRC13}, even allowing for a mixed
composition.  Thus a critical step required to fully understand
$X_{\rm max}$ observations is to identify and correct the deficiencies
in current shower models.  This is a strong motivation for upgrading
present-generation detectors to enable full understanding of the
hadronic interactions involved in air shower development.
Fortunately, the information which will be accessible in shower
observations -- including the correlation between $X_{\rm max}$ and
the ground signal in individual hybrid events~\cite{afICRC13}, the
comparison between $X_{\rm max}$ and $X_{\rm max}^{\mu}$ (the
atmospheric depth where muon production is maximum), the dependence of
ground signal on zenith angle, and other detailed shower observations
-- is so rich and multifaceted that it will enable composition and
particle physics to be disentangled~\cite{afICRC13}.

An additional intriguing twist in the present observational situation
is that the HiRes and TA results are consistent with a proton
dominated flux everywhere above the ankle~\cite{Abbasi:2009nf,
Tsunesada:2011mp}, although with present statistics the TA and Auger
elongation rates are consistent within
errors~\cite{TA-AugerER-ICRC13}.  Since the sources seen by the HiRes
and TA in the Northern hemisphere may not the same sources as seen by
the Auger Observatory in the South, the composition need not be the
same.  When TA statistics are sufficient to make a clear determination
as to whether the elongation rate observed in the North is the same as
recorded by the Auger observatory in the South, it will be of great
consequence for astrophysics even without knowing exactly how to
translate from elongation rate to composition.  If the composition
(elongation rate) in North and South are not the same, it will mean
{\it i)} that there are at least two source types, one accelerating
primarily protons and another accelerating a mixed composition, and
{\it ii)} that in at least one hemisphere, the UHECRs are produced
mainly by one or a small number of sources.  This is a strong motivation for increasing the aperture of
TA as described in Section \ref{sec:ta_upgrades}.

Another major result of the present generation of observatories is the
search for anisotropy in the distribution of arrival directions.
Around $10^{18}$ eV, Auger has provided a strong upper limit on the
dipole anisotropy~\cite{Abreu:2012lva,Auger:2012an} which is almost
sufficient to rule out a Galactic origin assuming these cosmic rays
are indeed predominantly protons and making reasonable assumptions
about the Galactic magnetic field (GMF).  When the TA and Auger data
are combined, the limit will be even stronger or a signal will be
found~\cite{delignyAugerICRC13}.

As the energy increases, evidence for anisotropy mounts. Auger has
reported a notable correlation of cosmic ray arrival directions with
nearby galaxies of the Veron-Cetty and Veron catalog of Active
Galactic Nuclei (AGN)~\cite{Cronin:2007zz}. With more data
accumulated, the central value of the correlation fraction has
decreased but the significance has remained at the 3-sigma
level~\cite{Abreu:2010ab,AugerICRC-HL12}.  The HiRes experiment did
not observe such a correlation~\cite{Abbasi:2008md}, but the most
recent results from TA ~\cite{AbuZayyad:2012hv,TinyakovTAanisoICRC13}
show a degree of correlation compatible with that seen by Auger in its
full data set, and with a similar pre-trial significance.
Furthermore, TA finds a significant correlation between the highest
energy events' pointing directions and the local large-scale structure
of the universe~\cite{TA-AugerAniso-ICRC13}.

While indications of anisotropy are becoming stronger, a completely
clear picture is thus far elusive, especially regarding the identity
of the sources themselves.  Perhaps a clear picture should not be
expected, given the possibility of multiple types of sources and that
fact that the composition in the South could be mixed or become heavy
at the highest energies, while at the same time the flux could be more
proton-dominated in the North.  Adding to the difficulty of comparing
correlation results of Northern and Southern hemisphere observatories
is the fact that the magnitude and directions of magnetic deflections
and the degree of multiple-imaging are expected to vary quite strongly
across the sky~\cite{fjr13}.  Fortunately astrophysical observations
and theoretical effort are rapidly improving GMF models~\cite{jf12},
so that the back-tracking to correct for deflections should become
feasible, to some extent, on the time-scale of the next generation of
experiments.

%% file: CF6/accel.tex

\subsubsection{Astrophysical Particle Acceleration and Gamma Ray Observations}
\label{sec:CF6-accel}

In the very-high-energy (VHE) band (defined here as gamma rays between
approximately 50 GeV and 100 TeV) one sees only non-thermal radiation,
mostly from the acceleration of charged particles by the most extreme
objects in the universe (some non-thermal radiation may be due to the
decay or annihilation of fundamental particles).  Within our Galaxy
these objects include supernova remnants, rapidly spinning neutron
stars, stellar mass black holes, and x-ray binary systems (composed of
a black hole or neutron star orbited by another star).  Extragalactic
objects that accelerate particles include supermassive black holes,
gamma-ray bursts, starburst galaxies, and galaxy clusters.  Many of
these objects have extreme gravitational and magnetic fields, which
may play a central role in the acceleration of particles in
relativistic jets.  By studying these sources we can discover the
sources of the Galactic and extragalactic cosmic rays, study physics
in extreme environments, and search for signatures of physics beyond
the Standard Model.

In the past decade we have discovered that particle acceleration is
ubiquitous in the universe.  As VHE telescopes have improved in
sensitivity we have found more objects and more classes of objects
where particle acceleration occurs.  From both gamma-ray and
cosmic-ray observations we know that somewhere in the universe there
are objects capable of accelerating fundamental particles to over
10$^{20}$ eV.  Understanding these extreme environments and how they
accelerate particles is of fundamental interest.  In addition, these
high-energy particle beams, visible from cosmologically interesting
distances, allow us to probe fundamental physics at scales and in ways
that are not possible in earth-bound particle accelerators.  In most
cases the astrophysical signatures of the particle acceleration
processes are encoded and multiplexed with signatures from physics
beyond the Standard Model. Thus, to extract information related to
Beyond the Standard Model Physics it is necessary to understand the
particle acceleration mechanisms and any imprints on the gamma-ray
energy spectrum that may arise from the local environment.

\paragraph{Galactic Sources of VHE Gamma Rays}
Cosmic rays were discovered over one hundred years ago
\cite{hess1912}.  Cosmic rays with energies below about 10$^{17}$ eV
are likely accelerated within the Galaxy.  Because protons and other
cosmic-ray nuclei are charged they bend in the Galactic magnetic field
and do not point to their origin.  Any site of cosmic rays should also
produce gamma rays.  Recent gamma-ray measurements by Fermi and
VERITAS have conclusively shown that supernovae are sites of
cosmic-ray acceleration \cite{ackermann2013a, acciarri2011}, solving a
century old problem.

The recent observations of an increasing positron:electron ratio above
100 MeV \cite{adriani2009, aguilar2013} may be attributed to the
annihilation of dark matter particles with mass $\sim$100 GeV.
However, before attributing this signature to the dark matter it is
important to understand any potential astrophysical backgrounds that
may contribute to (or even dominate) the positron fraction at high
energies.  Pulsar wind nebulae (PWN) are a potential source of
high-energy positrons that could mimic the signature from dark matter
decay.  Thus it is important to understand the energy spectrum of PWN
and the propagation of cosmic-ray electrons and positrons to
understand the origin of the PAMELA signal.  At VHE energies PWN
dominate the Galactic sources with over 30 discovered to date
\cite{tevcat}.  PWN accelerate electrons and positrons in equal
numbers at the termination shock of the pulsar wind.  These electrons
and positrons are then trapped by the magnetic field of the nebula and
produce gamma rays via inverse Compton interactions with the CMB and
low energy photons in the nebula.  As the nebula expands, the magnetic
field weakens and eventually the electrons and positrons are released
to propagate through the Galaxy.  Vela X-1 and Geminga are two nearby
pulsars that may contribute significantly to the local
positron/electron ratio.  Recent measurements of TeV gamma rays from
Geminga by the Milagro telescope \cite{abdo2009} have been used to
show that the local positron to electron ratio, based solely on
Geminga as a source, is consistent with the observations of PAMELA
\cite{yuksel2009}.  Data at higher energies will be needed to
distinguish a dark matter source (where one expects the ratio to fall
above the dark matter mass) from PWN.

\paragraph{Extragalactic Sources of VHE Gamma Rays}
The predominate class of extragalactic object detected in the VHE
energy band are active galactic nuclei (AGN), accretion powered
super-massive black holes with relativistic jets emitted along their
rotation axes.  VHE gamma rays have also been detect from starburst
galaxies \cite{acciarri2009}, presumably from cosmic-ray generation and
interactions in those galaxies.  Galaxy clusters and gamma-ray bursts
are other important potential sources of VHE gamma rays that so far
have eluded detection in this energy band.  Here we discuss these
sources and their implications for particle acceleration and
fundamental physics (details on the fundamental physics can be found
in other sections of this chapter).

VHE observations of galaxy clusters may provide the best sensitivity
to the annihilation or decay of dark matter.  However, there are
several astrophysical mechanisms by which galaxy clusters can produce
VHE gamma rays.  Galaxy clusters may contain active galaxies, known
sources of VHE gamma rays, structure formation shocks can accelerate
injected electrons and protons, and cosmic rays accelerated within the
cluster will produce gamma rays through the decay of neutral pions
when they interact with the intracluster medium.  While the
non-observation of VHE emission from a galaxy cluster can be used to
set stringent limits on the presence of WIMP-like dark matter, the
detection of a dark matter signal will rely on our understanding of
astrophysical production of gamma rays within galaxy clusters.

AGN have proved to be prolific sources of VHE gamma rays.  The
observed high-energy emission is strongly dependent on the viewing
angle with respect to the jet, as the bulk of the VHE gamma rays are
produced by processes within the jets.  AGN provide an excellent beam
of photons with which to perform fundamental physics studies.  If one
can understand the inherent spectrum of these objects the observed
spectrum at earth can be used to search for the existence of
axion-like particles.  Since active galaxies exhibit strong flaring
behavior on short timescales they can be used to time the relative
arrival time of photons of different energies to search for violations
of Lorentz invariance.  Since there are many active galaxies detected
over a large range of redshifts when can use the redshift dependence
of an observed signature to help disentangle the astrophysical
phenomena from the fundamental physics.  However, before extracting
such fundamental physics one must understand the acceleration
mechanism, the inherent spectrum of gamma rays, and the observed
source-to-source variations.

Currently there is much that is not understood about particle
acceleration in AGN.  The standard model of acceleration in jets is
based on the first-order Fermi mechanism, where particles are
accelerated via reflections from fast moving shocks.  However, the
importance of magnetic reconnection as an acceleration mechanism in
AGN is not understood and depends on the relative energy carried by
particles or an electromagnetic field.  The nature of the particles
accelerated by jets is also not known.  Most observations are
consistent with the acceleration of electrons and the subsequent
production of gamma rays via inverse Compton scattering from lower
energy photons.  However, hadronic models can not be ruled out and are
required if the ultra-high-energy cosmic rays are produced in AGN.

Gamma-ray bursts (GRBs) are the most violent objects in the universe,
emitting over 10$^{52}$ ergs in gamma rays.  There are believed to be
two classes of GRBs: the first whose progenitors are massive stars
undergoing core collapse supernovae and the other merger events of
neutrons stars or neutron star black hole systems.  In both cases it
is believed that accretion onto a black hole powers highly
relativistic jets where particles are accelerated to high energies.
In some GRBs a solar mass of material may be accelerated to bulk
velocities approaching the speed of light ($\Gamma \sim 1000$).  The
time duration of GRBs spans the range from a few milliseconds to 1000
seconds and in many cases fine substructure is present in the
gamma-ray arrival times.  Energy dependent time lags in the GRB onset
and the structure within a burst can be used to search for small
effects due to violation of Lorentz invariance.  As with AGN it is
necessary to understand the inherent time evolution of the GRB energy
spectrum to extract the fundamental physics.  To date the highest
energy gamma ray detected from a GRB had an energy at earth of 94 GeV
\cite{zhu2013}, which corresponds to an energy of 125 GeV at the GRB's
redshift of 0.34.  No convincing detection of a GRB has been made by a
ground-based VHE instrument, though the next generation of wide-field
instrument, HAWC, should have sufficient sensitivity to detect the
more energetic GRBs \cite{abeysekara2012}.

Detecting an axion signature in the energy spectrum of AGN requires
knowledge of the inherent AGN spectrum and the density of the
extragalactic background light (as a function of redshift).  To
attribute any observed energy dependent time delay of photons from
distant sources such as AGN or GRBs requires knowledge of the time
evolution of the gamma-ray spectra from these objects.  While the
current generation of VHE instruments has made enormous progress in
detecting new sources and classes of sources, the next generation of
instruments must have the precision and sensitivity to disentangle the
astrophysical processes from the fundamental physics processes.  This
will require excellent energy and angler resolution, multiple source
detections with a large span of redshifts, and multi-wavelength and
multi-messenger campaigns to truly understand the astrophysical
process of particle acceleration in these extreme objects.

%% file: CF6/neutrinos.tex

\subsection{Neutrino Physics from Astrophysics}
\label{sec:CF6-neutrinos}

\subsubsection{GZK Neutrinos}

Neutrinos result from the cascade of decays following cosmic ray
interactions with the microwave background.  The processes involved in
the production of these neutrinos are well-known low-energy particle
physics boosted by the large Lorentz factor of the cosmci ray-CMB
frame, and thus predictions of their fluxes are on a firm footing.
The flux of these neutrinos depends on the composition of the highest
energy cosmic rays and on the evolution of the cosmic ray sources
with redshift. Measurement of the GZK neutrino flux would help
disentangle the effenct of cosmc ray interactions, composition, and
source evolution, and provides a unique beam of high energy neutrinos
to probe neutrino cross sections at EeV energies.

\subsubsection{Supernova Burst Neutrinos}

\begin{figure}[ht]
	\centerline{\includegraphics[height=3in]{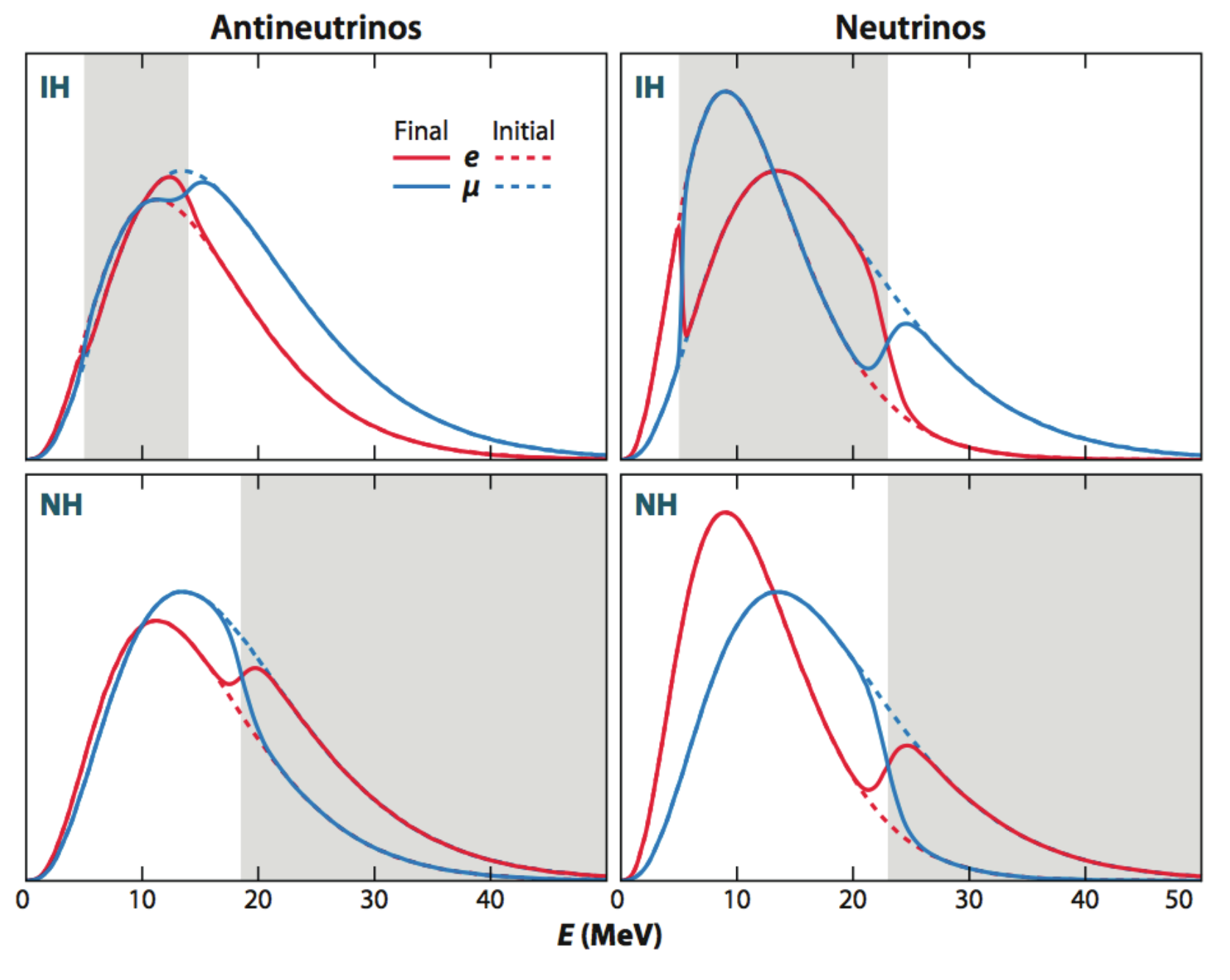}}
	\caption{The antineutrino (left panels) and neutrino energy
          spectra (right panels) for the normal hierarchy (bottom
          panels) and the inverted hierarchy (top panels).  In all 4
          figures the dashed lines show the observed spectra in the
          absence of collective oscillations and the solid lines the
          spectra after the effect of collective oscillations.  The
          figure is from \cite{dasgupta2009}}
	\label{fig:SNneutrinos}
\end{figure}

The measurement of the time evolution of the neutrino energy and
flavor spectrum from supernovae can revolutionize our understanding of
neutrino properties, supernova physics, and discover or tightly
constrain non-standard interactions.  Collective oscillations of
neutrinos leaving the neutron star surface imprint distinct signatures
on the time evolution of the neutrino spectrum that depend, in a
dramatic fashion, on the neutrino mass hierarchy and the mixing angle
$\theta_{13}$ \cite{duan2006, duan2007, duan2010}.  With $\theta_{13}$
now measured to be non-zero \cite{an2012} the measurement of the
electron neutrino energy spectrum from a supernova burst can be used
to determine the neutrino mass hierarchy.  Figure
\ref{fig:SNneutrinos} shows the resulting neutrino and antineutrino
energy spectra for the normal and inverted hierarchies, before and
after the spectral swapping from collective oscillations.  Note that
the change in the anti-neutrino energy spectrum (between the normal
and inverted hierarchies) is much less dramatic than for the neutrino
spectrum.  A liquid argon detector such as LBNE is most sensitive to
the electron neutrinos from a supernova burst measure the neutrino
through the reaction $\nu_e + ^{40}Ar \rightarrow e^{-} + ^{40}K^{*}$
and therefore best suited to measure the mass hierarchy of the
neutrinos with supernova neutrinos.  However, since the cosmic-ray
backgrounds at the surface are quite large, the detector must be
located underground to perform this physics.  In addition, with the
detection of ~1000 or more neutrinos from a supernova one could
temporally resolve the energy and flavor spectrum of the neutrinos and
glean information about the supernova process and the formation of the
remnant neutron star. For nearby Galactic supernovae, large detectors
such as IceCube complement low-background detectors by providing data
with high temporal resolution\cite{Abbasi:2011ss}.

\subsubsection{Atmospheric Neutrinos}
In addition to neutrinos from supernova bursts, nature provides a
steady source of neutrinos that can be used to determine the neutrino
mass hierarchy.  The atmospheric neutrinos generated by extensive air
showers travel through the earth.  A detector sensitive to these
neutrinos can measure neutrinos of with different energies that have
traversed a range of baselines, thus probing a large space of $L/E$.
The baselines available are approximately the diameter of the earth
($\sim$12700 km), in principle enabling sensitive searches for matter
effects and sensitivity to the neutrino mass hierarchy.  The Precision
IceCube Next Generation Upgrade (PINGU) experiment \cite{aartsen2013},
would provide an additional 20-40 strings (each with 60-100
photomultiplier tubes) to further ``infill" the DeepCore array
\cite{Collaboration:2011ym} within IceCube.  With a string spacing of
20-25 meters (compared to 73 m spacing within DeepCore and 125 m
spacing for IceCube), the energy threshold for neutrinos of would be
$\sim$5 GeV.

\begin{figure}[ht]
	\centerline{\includegraphics[height=3in]{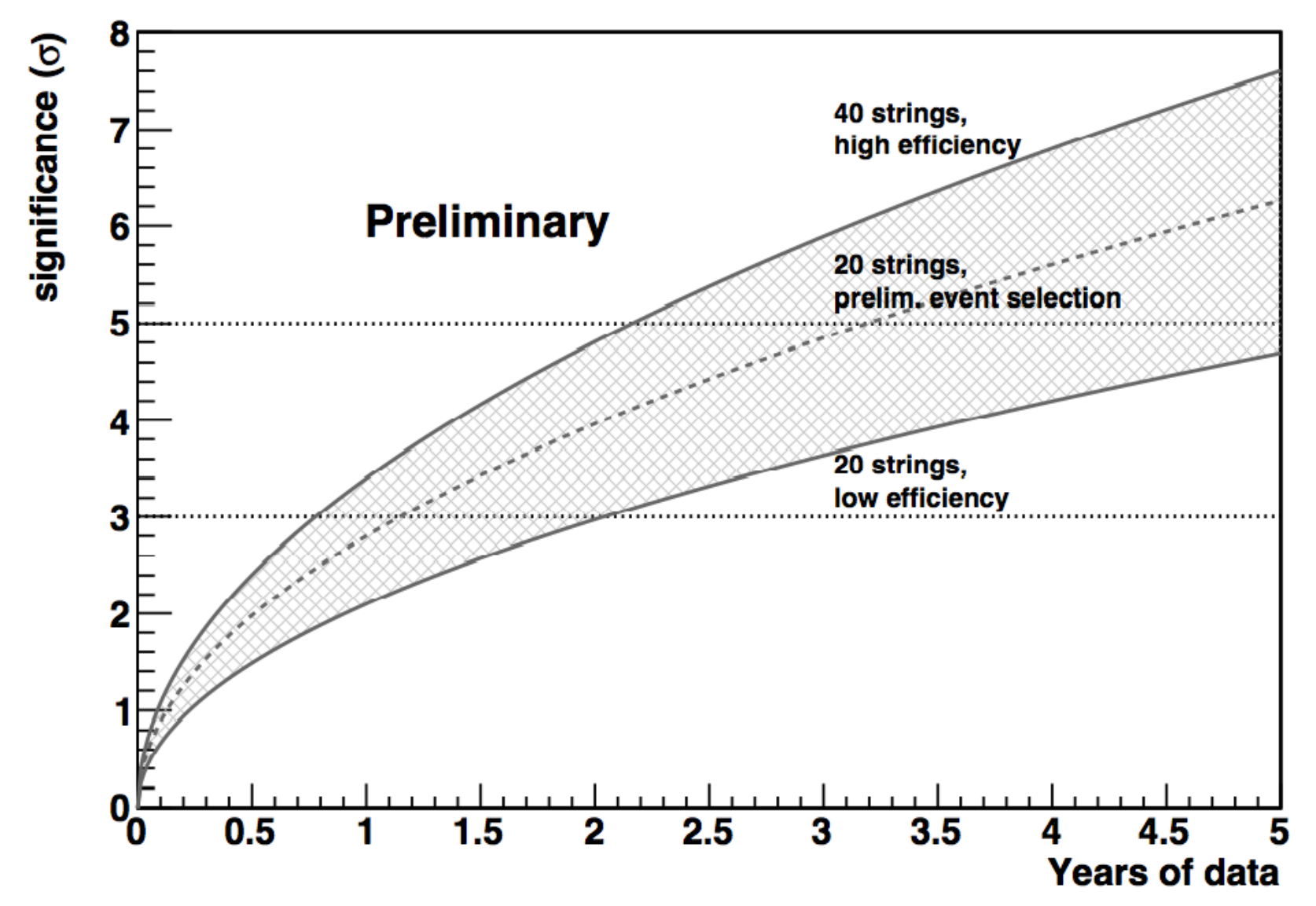}}
	\caption{Estimated significant for determination of the
          neutrino mass hierarchy with PINGU.  The top range is based
          on a 40 string detector with a high assumed signal
          efficiency in the final analysis.  The bottom curve uses a
          20 string detector and a lower assumed signal efficiency.
          The figure is from \cite{aartsen2013}}
	\label{fig:pingusense}
\end{figure}

Figure \ref{fig:pingusense} shows the expected statistical
significance on the determination of the neutrino mass hierarchy as a
function of time for PINGU.  The vertical band encompasses different
configurations (20-40 strings) as well as varying detector
efficiencies.  A 5 standard deviation measurement could be made in 2-5
years of data taking.  Since this is essentially a muon neutrino
disappearance experiment, comparing the surviving fraction of muon
neutrinos as a function of $L$ and $E$, proper knowledge of the
angular and energy resolutions of PINGU are key to understanding the
sensitivity to the neutrino mass hierarchy.  The studies presented in
\cite{aartsen2013} use algorithms developed for DeepCore to estimate
these resolutions and the curves in Figure \ref{fig:pingusense} are
based on those algorithms.

%% file: CF6/hadronicinteractions.tex
\subsubsection{Hadronic Interactions}

Ultra high energy cosmic ray measurements can be used to constrain
hadronlic interaction cross sections and mutliplicities at energies
far beyond those accessible using terrestrial accelerators.  While the
composition of the primary cosmic ray beam remains an open question,
even a small admixture of protons allows extraction of the proton-air
cross section from the tail of the distribution of the depth of shower
maximum $X_{max}$ toward large atmospheric depths. For example, Auger
data has been used to perform a measurement of the $p$-air
cross-section at $\sqrt{s} = 57~{\rm
TeV}$~\cite{Collaboration:2012wt}, a result which excludes some
hadronic models' extrapolations beyond LHC energies.

In addition, detailed analysis of the spectrum and anisotropy of
cosmic rays together with $X_{max}$ and other observables that are
diagnostics of composition and hadron interactions may ultimately
allow the composition and interactions of cosmic rays to be
disentangled.  Recent developments in this area are discussed below
in [cross reference to cosmic ray origins section].

%% file: CF6/neutrinointeractions.tex
\subsubsection{Neutrino Cross Sections at Extremely High Energies}

\begin{figure}[neutrino1]
\begin{center}
\includegraphics[width=0.5\hsize]{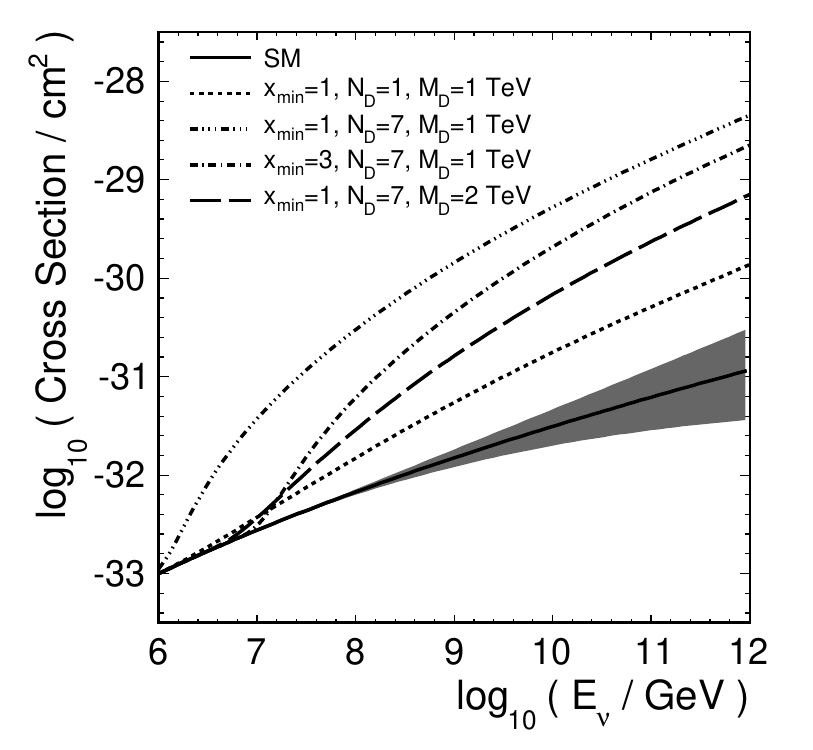}
\caption{The neutrino interaction cross-section for the standard model, plus several models that incorporate additional
dimensions.\cite{Connolly2011}}
\label{fig:neutrino1}
\end{center}
\end{figure}

Large neutrino telescopes can measure the neutrino-nucleon
cross-section by studying neutrino absorption in the
Earth \cite{Klein2013, hooper2002, Borriello2008}. At high energies, this
cross-section is sensitive to new physics. In particular, if there are
additional rolled-up dimensions, then the cross-section will increase
sharply at an energy corresponding to the inverse size of the extra
dimension(s). Figure \ref{fig:neutrino1} shows the neutrino-nucleon
cross-sections calculated for the Standard Model, plus several models
with additional dimensions\cite{Connolly2011}. Other types of new
physics can produce similar effects. For example, the presence of
leptoquarks could produce a similar increase in the
cross-section\cite{Romero2009}.

\begin{figure}[neutrino2]
\begin{center}
\includegraphics[width=0.5\hsize]{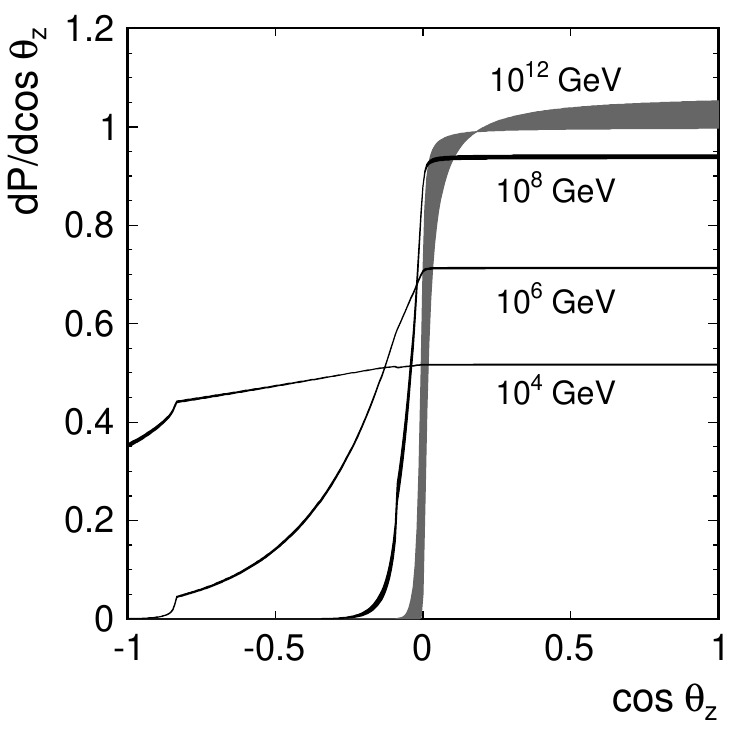}
\caption{The zenith angle distribution for high-energy neutrinos of different energies, for Standard Model cross sections.
At energies above $10^{17}$ eV, absorption in the Earth limits
the flux of upward-going neutrinos, except near the horizon.
\cite{Connolly2011}}
\label{fig:neutrino2}
\end{center}
\end{figure}

Neutrino absorption becomes an effective technique for measuring the
cross section at neutrino energies above about 50 TeV, the energy at
which absorption (assuming the Standard Model cross-sections) begins
to reduce the flux of vertically upward-going neutrinos, altering the
zenith angle distribution. Figure \ref{fig:neutrino2} shows the
expected zenith angle distribution for neutrinos with energies between
$10^{13}$ and $10^{21}$ eV. Lower energy neutrinos can be used as a
normalization, to check the angular acceptance of the detector, and to
calibrate for the small zenith angle dependence in the atmospheric
neutrino flux.

At neutrino energies much above $10^{17}$ eV, the cross sections
depend significantly on parton distributions at Bjorken $x$ and $Q^2$
values beyond the reach of HERA data, so extrapolations are required
to predict the cross sections. LHC data can be used to contrain the
parton distributions, but, even at current experiments like IceCube,
the neutrino energies are 100 times higher than are accessible at
accelerators. So, suprises are certainly possible, especially for new
physics with large couplings to the neutrino sector.

%% file: CF6/liv.tex

\subsubsection{Violation of Lorentz Invariance}
\label{sec:CF6-LIV}

The development of theories that merge general relativity and quantum mechanics - quantum gravity, typically lead to the violation of invariance principles that have been sacrosanct in physics.  The simple fact that quantum mechanics requires a minimum length scale (typically taken to be the Planck length (1.6$\times 10^{-35}$m or 1.2$\times10^{19}$ GeV) that is independent of reference frame, is in itself a violation of Lorentz invariance which is scale independent \cite{saslow1998}.  Probing physics at the Planck scale is difficult or impossible in earth-bound experiments, however nature has provided a mechanism by which we may probe certain aspects of Lorentz invariance violation to extraordinary precision.  In addition to quantum gravity, other motivations for considering violations of Lorentz invariance include the need for a high-energy cutoff to control divergences in quantum field theory \cite{rovelli2008}, and to develop a consistent theory of black holes \cite{solodukhin2011}.  One possible manifestation of the breaking of Lorentz invariance is a vacuum dispersion relation, an energy dependence to the velocity of light.  The energy dependence of the speed of light can be probed to high accuracy by measuring the time of high-energy, short pulses of light that have travelled cosmological distances.  Gamma-ray bursts and flares from active galaxies provide an excellent laboratory for such an investigation.  (For a review of other manifestations of a violation of Lorentz invariance see \cite{mattingly2005}.)

When considering a vacuum dispersion relation for electromagnetic radiation one typically performs a simple expansion with linear and
quadratic terms:
\begin{eqnarray}
\frac{v(p)}{c} = 1 + \zeta_1\left(\frac{p}{E_{QG}} \right) + \zeta_2\left( \frac{p}{E_{QG}} \right)^2
\end{eqnarray}
where $\zeta_n 's$ is either zero or one.  The linear and quadratic terms are often treated independently (i.e. one of the $\zeta_n^{'s}$ is zero.   The linear term violates CPT invariance \cite{colladay1997, colladay1998},while the quadratic term preserves CPT invariance.  The above dispersion relation then leads to an energy dependent delay \cite{amelino1998}:
\begin{eqnarray}
\Delta t \approx \left (\frac{\Delta E}{\zeta_n E_{QG}} \right)^{n} \frac{L}{c}
\end{eqnarray}
The dispersion relation may have a directional dependence \cite{kostelecky2009}, which argues for a sample of bursts and flares large enough to probe the dispersion relation in several directions.

The best limits on the linear term come from Fermi LAT observations of distant gamma-ray bursts and have probed to energies above the Planck scale
\cite{abdo2009}.  The best limits on the quadratic term (where energy reach is more important than distance) are derived from observations of flaring active galaxies by imaging atmospheric Cherenkov telescopes ($E_{QG}>6\times10^{11}$ GeV \cite{albert2008}.  Current limits on both the linear and quadratic terms are given in Table \ref{table:cf6-liv-1}.

\begin{table}[htdp]
\caption{Current lower bounds on the energy scale where Lorentz invariance is violated.  Limits for the linear(quadratic) term are given assuming that the quadratic(linear) term is zero. Adapted from \cite{bolmont2011}.  A superscript $l,q$ indicates that the limit applies to either the linear or quadratic term.}
\begin{center}
\begin{tabular}{| l | c | c | l |} \hline
Source & Experiment & Limit & Reference  \\ \hline \hline
Mrk 421 & Whipple & $E_{QG}^l$ $>$ 4$\times 10^{16}$ GeV & \cite{biller1999}  \\ \hline
PKS 2155-304 & H.E.S.S. & $E_{QG}^l$ $>$ 2.1$\times10^{18}$ GeV & \cite{abromowski2011}  \\ \hline
PKS 2155-304 & H.E.S.S. & $E_{QG}^q$ $>$ 6.4$\times10^{10}$ GeV & \cite{abromowski2011}  \\ \hline
GRB 090510 & Fermi GBM+LAT & $E_{QG}^l$ $>$ 1.5$\times10^{19}$ GeV & \cite{abdo2009}  \\ \hline
GRB 090510 & Fermi GBM+LAT & $E_{QG}^q$ $>$ 3$\times10^{10}$ GeV & \cite{abdo2009}  \\ \hline
\end{tabular}
\end{center}
\label{table:cf6-liv-1}
\end{table}%

Future experiments such as CTA and HAWC are expected to improve upon these limits and increase the sample size of bursts and flares to
probe any directional dependence to the vacuum  dispersion relation.  For example if HAWC observed a single gamma-ray burst at a redshift of 1
and could time the arrival of the high energy photons ($>$100 GeV) to one second, the limit on $E_{QG}$ from that single 
observation would be $4.9\times 10^{19}$ GeV for the linear term and $1.5\times 10^{11}$GeV for the quadratic term, roughly a factor of 2 higher than the best current limits \cite{nellen2013}.  The improvement from CTA is more dramatic.  CTA will have roughly an order of magnitude greater sensitivity than VERITAS or H.E.S.S.   Therefore CTA should improve the current limits on the quadratic term by a factor of 50 and be capable of probing to orders of magnitude above the Planck energy in the linear term \cite{doro2012}.   If AGN flares exhibit greater variability than current instruments can resolve, the limits from CTA will be even more sensitive.  

%% file: CF6/antimatter.tex
\subsubsection{Antiparticles and antinuclei}

Modern cosmological models strongly favor a universe in which an early baryon asymmetry develops followed by annihilation, leading to a
negligible abundance of antiparticles in the present-day universe.  Antiparticles and anti-nuclei can nevertheless be produced in energetic astrophysical processes and through decays.  

\paragraph{Electrons and Positrons}
Cosmic ray electrons are produced and accelerated in various astrophysical contexts and are considered primary cosmic rays, making up about 1\% of the total cosmic ray flux. No primary sources of cosmic ray positrons are known. However, positrons may also be injected by $\beta^+$ decays, from the magnetospheres of compact objects due to pair production, and as the result of primary cosmic ray interactions with the interstellar medium during the propagation. As a result, the number of cosmic ray positrons compared to the number of electrons is expected to decrease with energy. Observations by PAMELA \cite{PAMELApositron2009}, Fermi \cite{FERMIpositron2012}, and AMS-02 \cite{AMSpositron2013} have shown that the positron fraction instead rises with energies above 10 GeV. The high statistics positron fraction measurement by AMS-02 shows a steady increase up to 350 GeV, with a decreasing slope above 250 GeV. AMS-02 will expand the measurements to the TeV energy scale with 2\% energy resolution and unprecedented high statistics. These measurement could be a hint for a possible primary positron component as the kinematic characteristics of such a new primary process are most likely different from the production mechanisms of secondary positrons and could materialize in the form of an excess on top of the conventional diffuse positron spectrum. The interpretation will be tightly constrained by how the positron fraction continues at higher energies. Possible explanations span a wide range where dark matter particle self-annihilations or decay \cite{2013JCAP...05..003D,2013arXiv1304.1184K,2013arXiv1304.1482Y,2013arXiv1304.1483I,2013arXiv1304.2680K} and so far unknown astrophysical sources like nearby pulsars \cite{2013ApJ...772...18L,2013arXiv1303.0530F} are the most popular ones. Studying the anisotropy level of the positron flux helps to distinguish between diffuse and more concentrated sources. The AMS-02 limit on the dipole anisotropy of the positron
fraction is currently 3.6\% at the 95\% confidence level \cite{AMSpositron2013}. However, a $\sim$10 times lower level of anisotropy is expected from nearby pulsars \cite{2013ApJ...772...18L}.

\paragraph{Antiprotons}

The conventional astrophysical source of antiprotons is the production of antiproton-proton pairs in high energy collisions of cosmic rays with the interstellar medium. Low energy antiprotons can only be produced in high energy collisions, since sufficient energy must be available in the CM frame to allow antiprotons produced in the backward (anti-beam) direction to appear to be nearly at rest in the lab frame. These high energy collisions are rare due to the falling cosmic ray spectrum, so the flux of antiprotons below a GeV is kinematically suppressed.

The best measurements of the cosmic ray antiproton flux are from BESS \cite{PhysRevLett.88.051101,2008PhLB..670..103B} and PAMELA \cite{2010PhRvL.105l1101A}. Antiprotons are a powerful tool to constrain dark matter annihilation and decay models as the astrophysical production has only small uncertainties and the antiproton propagation is better under control than for positrons. However, dark matter annihilation (decay) fluxes from different channels (modes) are very similar in shape and, in addition,  also very similar in shape to the models for astrophysical production. This makes the interpretation challenging, but upcoming AMS-02 antiproton results have the potential to further constrain dark matter properties \cite{2013JCAP...04..015C}. The existing measurements do not require to introduce an additional primary antiproton component, which has to be taken into account for the interpretation of the positron fraction.

\paragraph{Antideuterons}

\begin{figure}
\centerline{\includegraphics[width=0.7\linewidth]{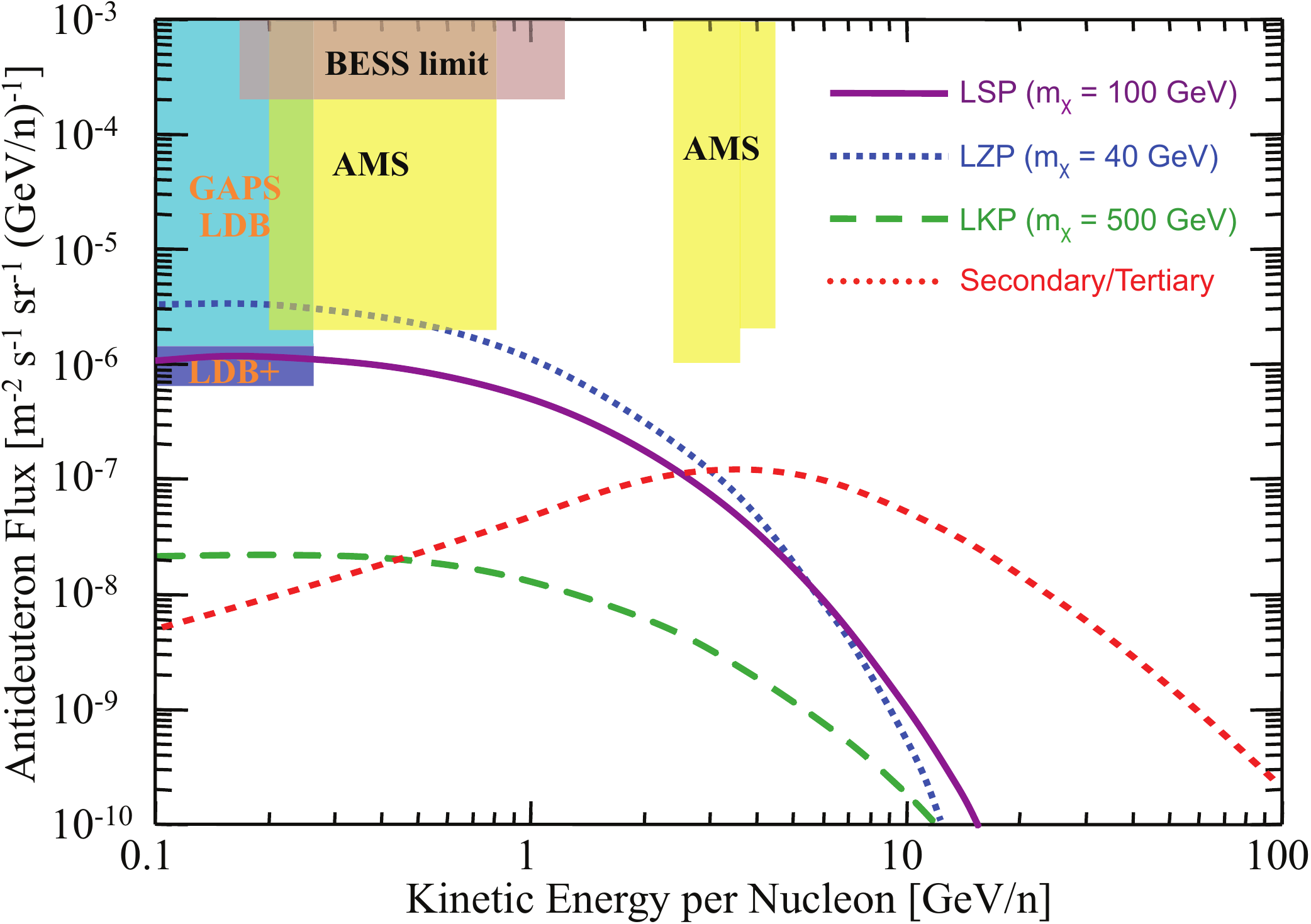}}
\caption{\label{fig:ams_gaps_dbar_sens}Predicted antideuteron fluxes from different dark matter models updated by more recent coalescence momentum value (purple, red, green lines) \cite{dbarbaer,ibarra20131} and secondary/tertiary background flux from cosmic ray interactions with the interstellar medium (blue line) \cite{ibarra20132}. Antideuteron limits from BESS \cite{bess} and sensitivities for the running AMS \cite{amsdbar,dbarsens} and the planned GAPS experiments \cite{gapssens} are also shown.}
\end{figure} 

Secondary antideuterons, like antiprotons, are produced when cosmic ray protons or antiprotons interact with the interstellar medium, but the production threshold for this reaction is higher for antideuterons than antiprotons. Collision kinematics also disfavor the formation of low energy antideuterons in these interactions. Moreover the steep energy spectrum of cosmic rays means there are fewer particles with sufficient energy to produce secondary antideuterons, and those that are produced will have relatively large kinetic energy. As a consequence, a low energy search for primary antideuterons has very low background. Figure~\ref{fig:ams_gaps_dbar_sens} shows the expected antideuteron flux from secondary and tertiary interactions as well as several dark matter models. The different boxes demonstrate the antideuteron flux limits of BESS \cite{bess} and the sensitivity reaches of AMS and GAPS \cite{amsdbar,dbarsens,gapssens}, which reach for the first time the sensitivity to probe predictions of well-motivated models. GAPS is a dedicated low energy antideuteron balloon experiment and had a successful prototype flight in 2012 \cite{2013arXiv1303.1615M,2013arXiv1307.3538V}. AMS and GAPS have mostly complementary kinetic energy ranges but also some overlap in the most interesting low energy region. Another very important virtue comes from the different detection techniques. AMS identifies particles by analyzing the event signatures of different subsequent subdetectors and a strong magnetic field and GAPS by slowing down antideuterons with kinetic energies below 300 MeV, creating exotic atoms inside the target material, and analyzing the decay structure. The combination of AMS and GAPS allows the study of both a large energy range and independent experimental confirmation, which is crucial for a rare event search like the hunt for cosmic-ray antideuterons.

%% file: CF6/pbh.tex

\subsubsection{Primordial Black Holes}
\label{sec:CF6-PBH}

 
Primordial density fluctuations can lead to the formation of black holes, the typical black hole mass will be of order the horizon 
mass (or smaller) at the time of formation ($M\approx10^{15}(t/10^-{23}s)$g)\cite{zeldovich1967}.  Therefore the mass spectrum of PBHs can span the range from supermassive black holes to Planck mass black holes.  Hawking demonstrated that black holes radiate energy and have a temperature that is proportional to the inverse of their mass ($T\approx 1/M_{13}$ GeV, where $M_{13}$ is the black hole mass in units of $10^{13}$g {\cite{hawking1974}.  Primordial black holes with an initial mass of less than $10^{15}$g would have evaporated completely.   Such black holes would have been formed in the first $10^{-23}$seconds after the Big Bang.  The cosmological consequences of primordial black holes are many, their existence (or absence) and mass spectrum probe the density perturbations at early times and very small scales \cite{kim1999, carr2010}. They can seed dark matter clumping, forming ultra compact minihalos and primordial stellar mass size black holes (formed during the QCD phase transition about 10 microseconds after the Big Bang \cite{jedamzik1997}) may make up the dark matter \cite{lacki2010, hawkins2011}.  

The luminosity of a black hole is inversely proportional to the square of the black hole mass and the proportionality constant is related to the number of available degrees of freedom.  The emission mechanism can be thought of as the creation of particle anti-particle pair creation at the event horizon, with one of the particles being absorbed and the other emitted.  All fundamental particles will be produced at the event horizon if their mass is less than the black hole temperature.  Therefore the time evolution of the emission spectrum of a PBH is sensitive to physics at energy scales not attainable on earth.  The observed emission spectrum can be found by convolving the emitted particle spectrum with particle fragmentation functions to derive the gamma-ray and neutrino luminosity functions \cite{halzen1991}.  Black holes with an initial mass of roughly $5\times 10^{14}$g would now be in their final stages of evaporation, emitting GeV-TeV particles.  

The discovery of a primordial black hole would have an enormous impact on science - informing us of the existence of particle states at extremely high energies (well above those available at the LHC) and the spectrum of the primordial density fluctuations at extremely small scales (roughly 30 orders of magnitude smaller than the scales probed by the CMB measurements).

Experimentally there are several ways to search for the existence of PBHs.  Indirect techniques rely on detecting the cumulative emission of particles (typically anti-protons or ~100 MeV gamma rays) by black holes that have or are in the process of evaporating.  Direct detection techniques rely on the detection of a black hole in the final stages of evaporation.  Understanding the relationship between the two techniques is complex as it depends upon the mass spectrum of primordial black holes and their clustering (direct techniques are sensitive to relatively local black hole evaporation).  While the indirect methods have essentially reached the limits of their sensitivity (the actual measured anti-proton and diffuse 100 MeV gamma-ray flux is used to set limits on the number of primordial black holes that could have evaporated in the past), direct techniques hold the promise of significantly improving upon the current limits.

The diffuse gamma-ray flux measured by EGRET has been used to set an upper limit of $\Omega_{pbh} < (5.1\pm 1.3)\times 10^{-9} h_0^{-2}$ \cite{carr1998} on the
contribution to the mass density of the universe from primordial black holes of mass less than 10$^{15}$ g.   If PBHs cluster in our Galaxy as ordinary matter the offset of the Sun from th Galactic Center would lead to an anisotropy in the fraction of the diffuse gamma-ray background that is due to emission from PBHs.  Wright \cite{wright1996} found such an anisotropy in the gamma-ray background measured by EGRET, however at a level much smaller than expected if PBHs clump as luminous matter.  From the measured anisotropy one can set an upper limit on the rate density of evaporating PBHs in the Galactic halo of $< 0.42 $pc$^{-3}$yr$^{-1}$ \cite{wright1996}.

The evaporation of PBHs would produce equal numbers of protons and antiprotons (and similarly for anti deuterons and deuterons).  A PBH signature would manifest itself as an increase in the antiproton to proton fraction at low energies (below $\sim$1 GeV, where the effects of solar modulation are important).  The absence of such an increase and the measured value of the antiproton flux by the BESS-Polar II instrument sets an upper limit to the local (within a $\sim$few kpc \cite{maki1996}) rate density of evaporating PBHs of $1.2\times10^{-3}$pc$^{-3}$yr$^{-1}$ \cite{abe2012}.  Anti-deuteron data from AMS and or GAPS could improve upon this limit \cite{barrau2003}.

Experiments sensitive to the final stages of the PBH evaporation process, IACTS and EAS arrays, probe the very local distribution of PBHs with mass near $5\times10^{14}$ g.  Figure \ref{fig:pbh-limit} shows the current limits and the expected sensitivity of future instruments (HAWC and CTA).  Note that the CTA limit is an estimate based upon the current limit established by the H.E.S.S. collaboration, the increased sensitivity of CTA compared to H.E.S.S. (10$x$ sensitivity allows one to probe a volume $10^{3/2}$ greater), and the larger field-of-view of CTA relative to H.E.S.S.  CTA could improve upon current limits by 2 orders of magnitude.

\begin{figure}
\centerline{\includegraphics[height=3.in]{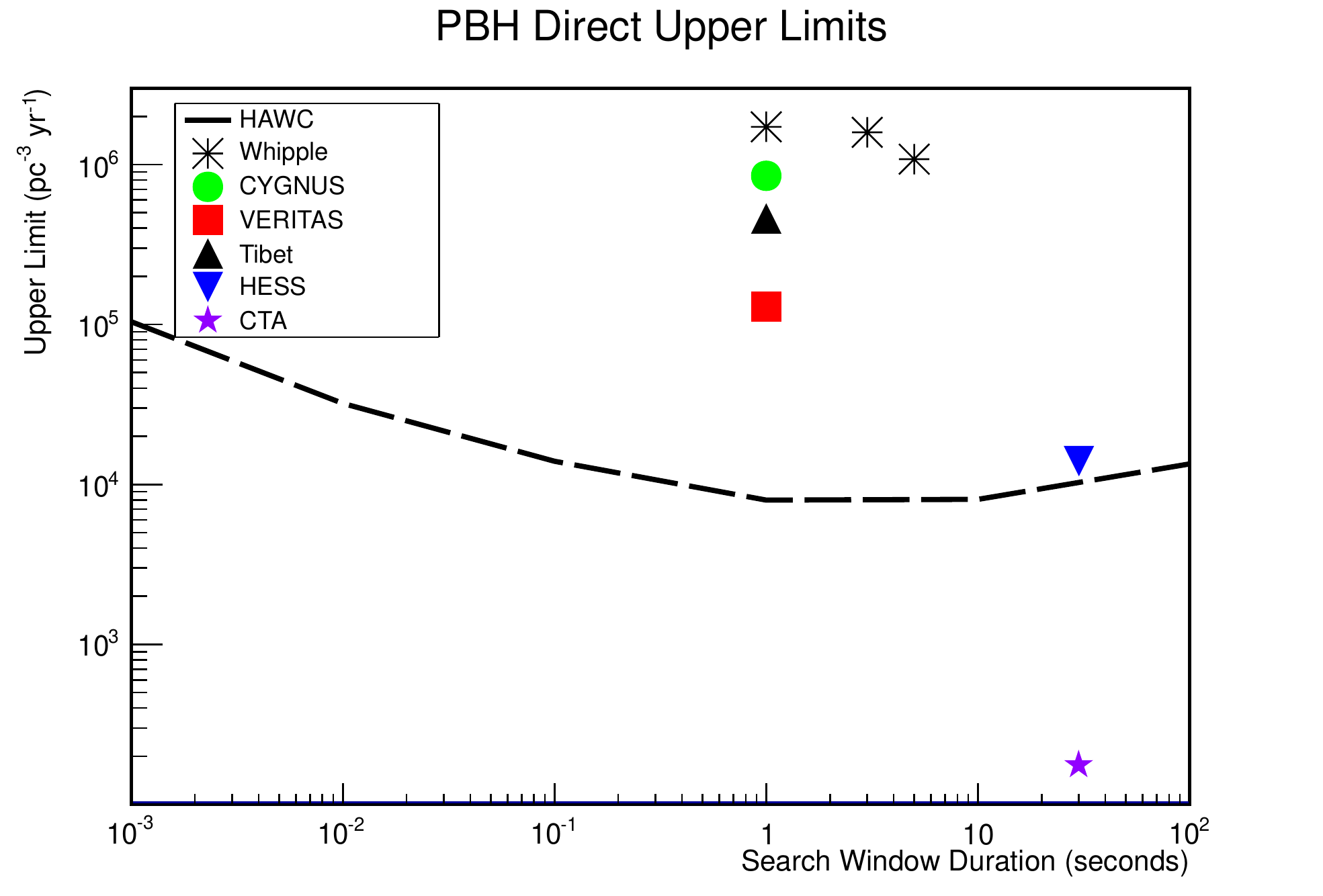}}
\caption{Current and potential limits on the local rate-density of evaporating primordial black holes.  The sensitivity of CTA has been estimated see text for details.  References for figure: HAWC \cite{ukwatta2013}, Whipple \cite{Linton:2006yu}, VERITAS \cite{tesic2012}, CYGNUS \cite{Alexandreas:1993zx}, Tibet \cite{amenomori1995}, and H.E.S.S. \cite{glicenstein2013}. }
\label{fig:pbh-limit}
\end{figure}

%% file: CF6/alp.tex

\subsubsection{The Extragalactic Background Light and the Search for Axion-like Particles}
\label{sec:CF6-alp}

\begin{figure}[h]
	\centerline{\includegraphics[height=3in]{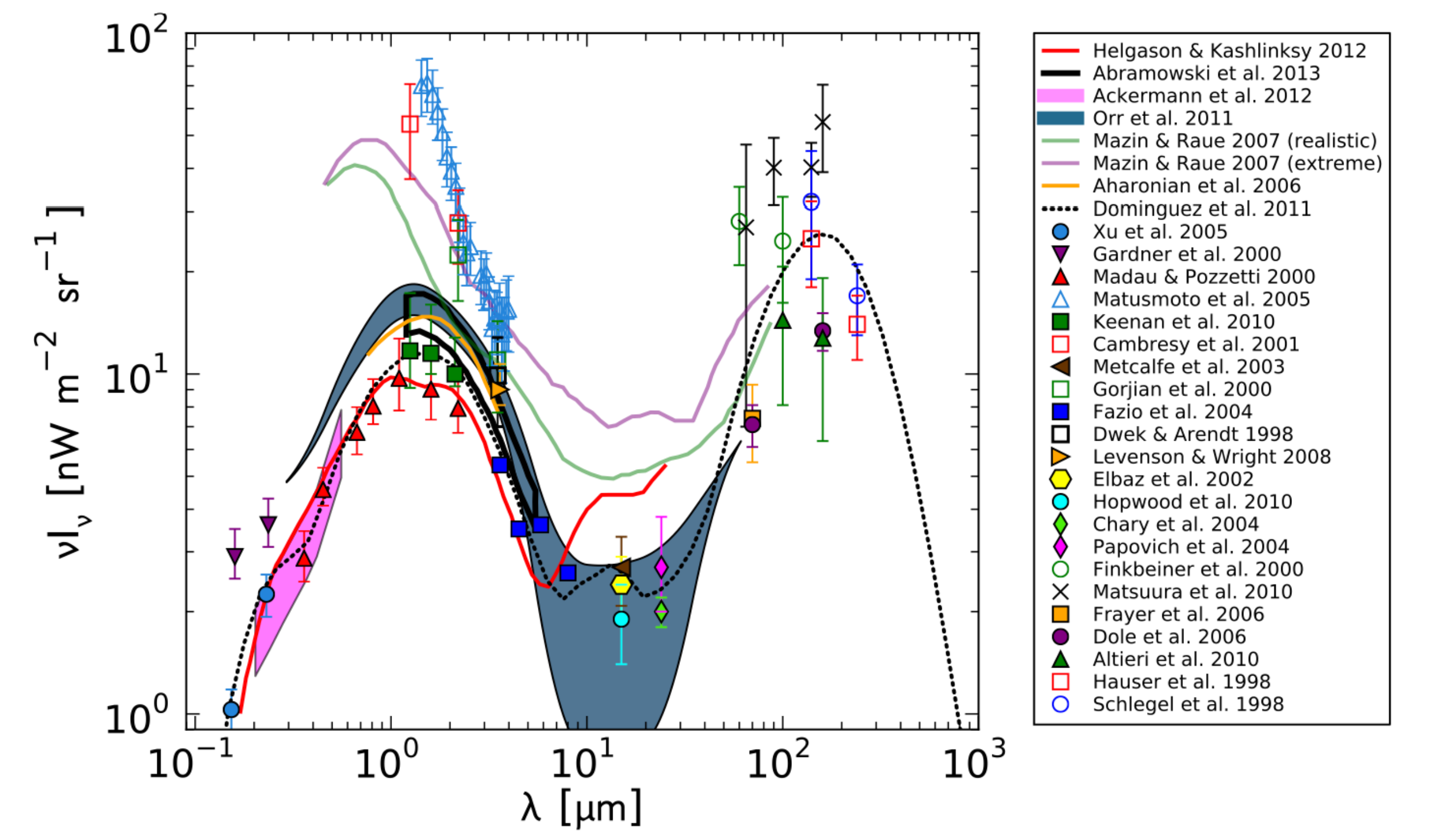}}
	\caption{A summary of our current knowledge of the
          extragalactic background light intensity as a function of
          wavelength.  Direct measurements are shown with open symbol,
          limits from IACT blazar observations are given by the legend
          listings 2-7, legend 1 is a model based on galaxy luminosity
          functions and the dashed black line is the model of
          Dominguez et al. \cite{dominguez2011}.  Other legend
          references can be found in \cite{dwek2013}.  This figure was
          taken from \cite{krennrich2013}.}
	\label{fig:ebl}
\end{figure}

The extragalactic background light (EBL) is the sum over the history
of the universe of the ultraviolet (UV) through far infrared (IR)
photon fields and their evolution.  The light sources are
predominantly stars and starlight reradiated in the infrared (IR) band
by dust.  The EBL thus contains an imprint of the past history of the
universe including galaxy and star formation and any other sources of
radiation.  In principle it is possible to calculate the EBL spectrum
from first principles - initial mass functions of stars, stellar
evolution theory, and the effect of dust.  Alternatively one can sum
the radiation fields from observed galaxy counts.  A direct
measurement of the EBL could then be compared to these estimates.  If
the measured EBL is higher than expectations this would be a signature
of the injection of a new radiation source, for example the decay of
weakly interacting particles or an early population of massive stars.
If the measured EBL (as measured by the resultant opacity of the
universe at VHE energies, see below) is lower than expected, this
would be a strong indication that the VHE photons are mixing with an
axion-like particle via interactions with intergalactic magnetic
fields or that distant high-energy gamma rays are a by-product of the
production of the ultra-high-energy cosmic rays.  This uncertainty is
borne out by calculations based on observational uncertainties in
determinations of galaxy luminosity densities\cite{Stecker:2012ta}.
Just as the CMB is responsible for the absorption of ultra-high-energy
protons \cite{Greisen:1966jv,Zatsepin:1966jv} and photons above 100
TeV, the EBL interacts with gamma rays with energies between $\sim$100
GeV and $\sim$100 TeV through the resonant production of
electron-positron pairs, resulting in an energy and distance dependent
attenuation of VHE gamma rays from distant sources.  Direct
measurements of the EBL are difficult due to sources of foreground
light.  If one understands the inherent energy spectra of AGN then the
observed spectra at earth in the VHE regime can be used to measure the
spectrum of the EBL.  To date the best upper bounds on the EBL
spectral energy distribution (SED) are derived from observations of
AGN spectra.  These limits are near the lower bounds established by
the contribution to the EBL from galaxies.  Figure \ref{fig:ebl} shows
the current limits and measurements of the spectral energy
distribution of the EBL \cite{krennrich2013}.

As can be seen from the
figure there is a large uncertainty in our knowledge of the SED of the
EBL.  The sensitivity of CTA to distant AGN should dramatically
improve our knowledge of the EBL and therefore our understanding of
the history of star and galaxy formation in the universe.

\paragraph{Axion-like Particles}
Axions were first postulated in \cite{peccei1997} to solve the strong
CP problem (the fact that CP symmetry is apparently conserved in QCD).
Such axions couple to the electromagnetic field through a term in the
Lagrangian of the form $g_{a\gamma}\mathbf{E \cdot B}a$, where $a$ is
the axion field and $g_{a\gamma}$ is the axion photon coupling, which
for traditional axions is inversely proportional to the axion mass.
Axion-like particles (ALPs) have a similar coupling to the
electromagnetic field, however they do not have the same relationship
between coupling strength and mass.  Axions and axion-like particles are a leading dark matter
candidate and searches for them are discussed in CF-3 of this
document.  A common feature of axion/ALP searches is exploitation of the
axion/ALP photon coupling in the presence of a large magnetic field.  In
the cosmic arena gamma rays emitted by distant sources such as AGN can
mix with ALPs via the intergalactic magnetic fields (or stronger
local fields near the acceleration region).  A fraction of these
ALPs then reconvert to gamma rays before reaching the earth.  While
this would lead to a decrease in the gamma-ray flux reaching the earth
in the absence of the EBL there can be an increased gamma ray flux at
earth compared to the EBL absorbed expectations.  The spectrum
measured at earth will depend upon the intrinsic AGN spectrum, the EBL
spectral energy distribution, and the intergalactic magnetic fields.
These measurements can provide the most sensitive searches for
axion-like particles with extremely low masses.

We are now entering an era where there significant number of AGN have
been observed over a broad energy range from 100 MeV - few TeV.  A
detailed analysis of 50 AGN spectra (out to a redshift of 0.5)
measured in the TeV region has found evidence for a suppression of EBL
absorption at the 4.2 standard deviation level \cite{horns2012}.  The
measured spectra are consistent with expectations based on photon-ALP
mixing, with an IGMF of 1 nG and a photon-ALP coupling strength of
$g_{a\gamma}=10^{-11}$ GeV$^-{1}$ (below the upper bound from CAST
\cite{ferrer-ribas2012}).  Though given the large range of
IGMF/$g_{a\gamma}$ parameter space available it is difficult to draw
strong conclusions based upon this analysis.

Another approach is to search for irregularities in AGN spectra,
expected if there is a photon-ALP coupling due to magnetic field
irregularities in the IGMF.  The H.E.S.S. collaboration, finding no
such irregularities has set limits on the photon-alp coupling of
$2\times 10^{-11}$ GeV$^{-1}$ in the mass range of near 20 neV
\cite{wouters2013}.  This limit is the most restrictive in this mass
range.

\begin{figure}[h]
	\centerline{\includegraphics[height=3in]{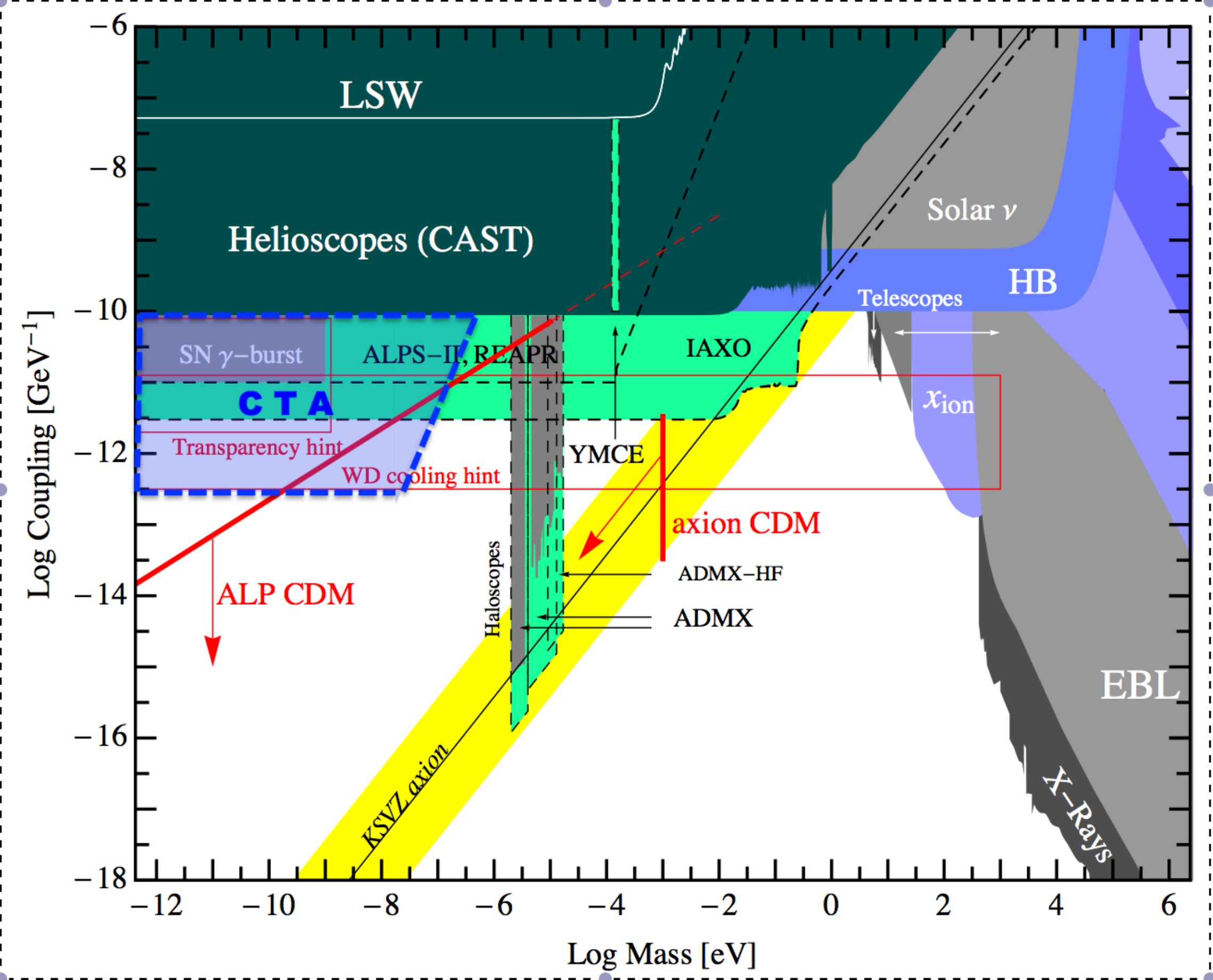}}
	\caption{Current limits and expected sensitivity of axion
          searches.  The potential sensitivity of CTA to axion-like
          particles is delimited by the dashed blue line.  Note a limit on the coupling strength of ~1.3-2.3$\times10^{-12}$ GeV$^{-1}$ 
          (for axions with mass less than $\sim$16meV) from the duration of the neutrino burst from SN1987A is not shown in the figure.  
          See text for an explanation.}
	\label{fig:alpspace}
\end{figure}

At present the situation is in undecided and it should be noted that
if AGN produce ultra-high-energy protons, interactions of these
protons with the CMB will produce secondary gamma rays that can also
lead to an enhanced flux of VHE gamma rays (over that expected with
EBL absorption) \cite{Essey:2009zg,Essey:2009ju,Essey:2010er,Essey:2010nd,
Essey:2011wv,Murase:2011cy,Razzaque:2011jc,Prosekin:2012ne,
Aharonian:2012fu,Zheng:2013lza,Kalashev:2013vba,Inoue:2013vpa}.  The
observation of a single flare at energies above $\sim$1 TeV from a
distant AGN ($z>0.15)$ could negate the UHECR interpretation.  What is
needed are more measurements spanning a larger redshift range, with
greater sensitivity and energy resolution from 100 GeV to 10 TeV.  CTA
should detect at least an order of magnitude more AGN and HAWC can
monitor every AGN in its field of view for flaring activity - for
detailed followup observations by CTA.  Figure \ref{fig:alpspace}
shows the current limits on $g_{a\gamma}$ as a function of axion mass
along with an estimate of the expected improvement in sensitivity from
CTA \cite{sanchez-conde2013}.  The line marked "WD cooling hint" comes from the observation that the cooling rate white dwarf stars 
is faster than expected, possibly indicating another energy loss mechanism (such as ALPs) \cite{Kuster:2008zz}.
Not shown in the figure is a limit on the coupling for axion masses less than $\sim$16 meV around
$1.3\times10^{-12}$ to $2.2\times10^{-12}$ from the duration of the neutrino burst from SN1987a (an additional energy loss mechanism would have shortened the duration of the neutrino burst) \cite{Kuster:2008zz}.  The expected sensitivity of CTA could probe the region of the white dwarf cooling hint, which is below the bound from SN1987a.

\subsubsection{Intergalactic Magnetic Fields}
Intergalactic magnetic fields (IGMFs) offer a new window on the
 early-universe cosmology and new physics.   The magnetic fields deep in
 the voids between galaxies, where no significant star formation or gas
 convection could take place, are the closest approximation to
 primordial seed magnetic fields left over from the Big Bang.  These
 fields are vestiges of some non-thermal events in the early
 cosmological history.  They could have been produced at the end of
 inflation from inflaton dynamics, or in the course of a cosmological
 phase transition, or due to some other processes in the early universe
 that involve new physics~\cite{Kandus:2010nw}.  The strength of IGMFs
 is still poorly understood~\cite{Kronberg:1993vk,Durrer:2013pga}. Until
 recently, only the upper limits of $10^{-9}$~G were inferred from the
 observational data~\cite{Barrow:1997mj}.  Theoretical models assuming
 the dynamo origin of galactic magnetic fields require primordial seed
 fields of $B>10^{-30}$~G~\cite{Davis:1999bt}, which can be considered a
 theoretical lower limit.

 With the advent of a new generation of gamma-ray instruments, one can
 measure the values of IGMFs deep in the voids, along the line of sight
 to distant gamma-ray sources, such as blazars.  One can measure IGMFs
 using time delays~\cite{Plaga:1994hq}, or by searching extended halos
 around the point objects~\cite{Aharonian:1993vz,Ando:2010rb}.
 Recent analyses of spectra of distant blazars produced both lower and
 upper bounds:  0.01 fG $<$ B $<$ 30 fG on Mpc scales~\cite{Essey:2010nd}.
 The narrowing of the range of primordial fields to a vicinity of a
 femtogauss~\cite{Essey:2010nd,Essey:2010er,Neronov:2013lta,Essey:2009zg,Essey:2009ju}
 has already stimulated a number of studies in primordial
 magnetogenesis~\cite{Kandus:2010nw,Maeda:2011uq,Demozzi:2012wh,Barrow:2012ty,
 Fujita:2012rb,Beck:2012cs,Kahniashvili:2012vt,Barrow:2012ax,Ringeval:2013hfa}.

 Future observations of distant blazars with CTA will allow mapping out
 of IGMFs, as well as a determination  of both the average strengths and
 the correlation length distributions of these fields. These observations can
 also measure the presence of helical magnetic fields~\cite{Tashiro:2013bxa},
 predicted in scenarios in which cosmic baryogenesis
 and magneto-genesis occur concurrently during a cosmological phase
 transition~\cite{Vachaspati:1991nm,Joyce:1997uy,Cornwall:1997ms,Vachaspati:2001nb,Tashiro:2012mf}.
 In addition to providing a clear signal for a primordial origin of the IGMF,
 helicity is also an important factor in the evolution of magnetic fields
 enabling their growth to astrophysically relevant scales at the present epoch.

This will offer a new exciting possibility to test models of inflation and new physics
via a new and unique window on the early universe.

%% file: CF6/monopole.tex
\subsubsection{Magnetic Monopoles}

The possibility of magnetic monopoles goes back at at least to 1931
\cite{Dirac}.  It is theoretically attractive, as it could explain the
observed quantization of electrical charge.  This argument also gives
the 'natural' size of the magnetic charge.  Monopoles appear naturally
in many grand unified theories (GUTs).  In most of these theories,
monopoles have masses comparable to the unification scale, and so are
far beyond the reach of current or planned accelerators.  However,
they might have been produced in the early universe
\cite{Kibble:1976sj}; since they do not decay, they may still be
present.

Direct and indirect techniques have been used to search for monpoles.
The indirect searches rely on how monpoles effect on various
astrophysical phenomena.  For example, the Parker bound
\cite{Turner:1982ag} is based on the fact that a sufficient density of
magnetic monopoles would short circuit the galaxys magnetic fields.
From the existence of these fields, a flux limit is set at roughly one
monopole per football field per year; this is roughly the upper limit
for direct searches to be useful.

The required technique for direct searches depends on the monpole
velocity.  Heavy (GUTs scale) monopoles are expected to be slow, with
velocities of order $10^{-4}$ - $10^{-3}$ $c$.  The most sensitive
search for monpoles with this velocity was by the MACRO experiment,
which set limits a factor of several below the Parker bound
\cite{Ambrosio:2002qq}.  Newer experiments have set lower limits on
non-relativistic monpooles, but only if they catalyze proton decay
\cite{IceCubeMP}.  The catalysis of proton decay is not unexpected in
GUTs theories, but the details are heavily theory-dependent.  These
experiments set two-dimensional limits, in terms of monopole flux
catalysis cross-section.  Assumptions are also required about the
final state(s) of the decaying proton.

If monopoles have masses below the GUTs scale, than they could be
accelerated to relativistic velocities. Then, they would emit
Cherenkov radiation, and could be detector by neutrino telescopes.
This signature has some similarities to that for nuclearites.  Several
neutrino detectors have searched for relativistic monopoles, with
negative results
\cite{Abbasi:2012eda,Detrixhe:2010xi,Pavalas:2013ima,Antipin:2007zz}.
These experiments mostly observe monopoles underground, so one must
account for the energy loss before the monopole reaches the detector.
They set limits ranging from of order 1/10 of the Parker limit (for
slightly relativistic monpoles, observed in optical detectors, to
limits several orders of magnitude tighter, for ultra-relativistic
monpoles observed with radio detectors.

A future dedicated experiment for non-relativistic, non-catalyzing
significantly larger than MACRO would be quite expensive, and, absent
new findings, is probably not worthwhile.  However, optical and radio
Cherenkov detectors will continue to set tighter limits.  Although the
searches are quite speculative, a positive detection would drastically
change our view of particle physics.

%% file: CF6/qballs.tex
\subsubsection{Q Balls}
\label{sec:CF6-QBall}

Understanding the origin of the matter anti-matter asymmetry in the
universe is one of the most fundamental problems facing modern
physics.  Though there are several viable explanations, experimental
evidence has been lacking.  Leptogenesis is being probed by
experiments such as the Long Baseline Neutrino Experiment (LBNE),
which is searching for CP violation in the neutrino sector, and
neutrinoless double beta decay experiments, sensitive to the Majorana
nature of the neutrino.  Other sources of baryonic CP violation are
being probed with $\mu \rightarrow e \gamma$ experiments and searches
for neutron and electron dipole moments.  An alternate scenario for
baryogensis, the Affleck-Dine mechanism \cite{affleck1985}, active at
the end of the inflationary epoch, can naturally generate a large
baryon asymmetry - if supersymmetry is the correct theory of
unification.  A prediction of this theory is the formation of Bose
condensates of the scaler field.  The condensates of squarks would
form Q-balls, potentially stable states (Q-balls with masses larger
than about 10$^{15}$ GeV are stable) with large baryon number.  Stable
Q-balls are candidates for the dark matter in the universe.
Reasonable values of Q-ball charge and the SUSY breaking scale put the
detection of Q-balls within the reach of planned high-energy
gamma-ray, cosmic-ray, and neutrino telescopes.

S. Coleman \cite{coleman1985} first realized that supersymmetric
theories have non-topological stable solutions which he called
"Q-balls".  A review of Q-balls and baryogenesis may be found
in \cite{dine2004}.  Here we give some of the important parameters for
Q-balls:

\begin{eqnarray}
M_Q = \frac{4\pi\sqrt(2)}{3} M_S Q^{3/4} ~GeV \\ \sigma = \pi
 R^{2}_{Q}
 = \frac{\pi}{2} \frac{Q^{1/2}}{M_{S}^{2}} \approx \frac{10}{M_{S}^{2}}\sqrt{Q/10^{14}}
 ~mbarns \\
\end{eqnarray}

where $M_S$ is the SUSY breaking scale in TeV and $Q$ is the baryon
number (charge) of the Q-ball.  Note that the cross section is purely
geometrical and quite large.  Since the mass grows slower than the
charge Q-balls will be stable, when it is energetically impossible for
them to decay into an equivalent number of protons, which occurs at
$Q>5.0\times 10^{14}(M_S/TeV)^4$ \cite{arafune2000}.  Because of their
high mass, the expected flux of Q-balls at earth would be low (if they
comprise the dark matter in the universe):

\begin{eqnarray}
F \approx \frac{\rho_{DM}v}{4\pi M_Q} \approx
7.2\times10^{5} \left(\frac{GeV}{M_Q}\right) cm^{-2}sec^{-1}sr^{-1}
~\cite{arafune2000}.
\end{eqnarray}
where we have assumed that $\rho_{DM}=0.3$ GeV/cm$^3$ and $v=300$km
sec$^{-1}$, the virial velocity.  For stable Q-balls and interesting
values of the SUSY breaking scale Q-ball fluxes of $\sim$10$^{-16}$ or
less may be expected.  Such a low flux requires the large effective
areas that are now common in cosmic-ray, gamma-ray, and neutrino
experiments.

The interaction of a Q-ball with ordinary matter proceeds would leave
a spectacular signal in the appropriate detector.  As a Q-ball
interacts with a proton, the proton is absorbed (increasing the baryon
number of the Q-ball) and an anti-proton is emitted.  Therefore at
each interaction with a nucleus the energy released into the detector
will be approximately 1 GeV/nucleon, most of which will be in the form
of pions.

\begin{figure}
\centerline{\includegraphics[height=3.in]{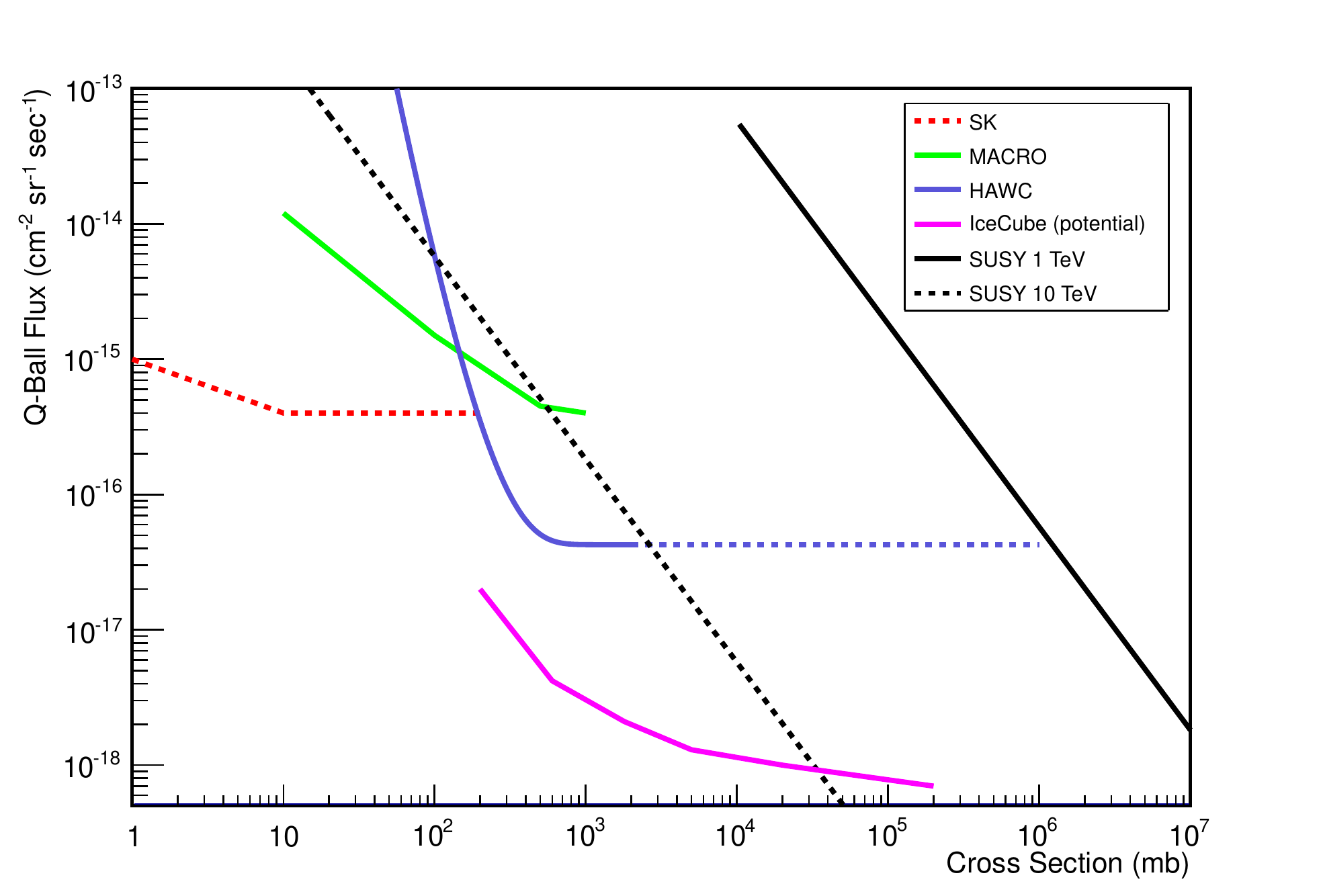}}
\caption{Current and potential limits on the flux of Q-balls as a function cross section from current and future neutrino, gamma-ray, and cosmic-ray instruments.  The diagonal lines are possible values of flux as a function of cross section for different values of the SUSY breaking scale assuming that the dark matter is entirely composed of Q-balls.  The IceCube potential limit is taken from an unpublished limit on monopoles with a velocity of 0.001$c$, a typical expected velocity for Q-balls.  References for the figure: SuperKamiokande \cite{Takenaga2007} , MACRO \cite{ambrosio2002}, HAWC \cite{karn2013}, IceCube \cite{deyoung2013}. }
\label{fig:qball-limit}
\end{figure}

Currently the best limits on Q-balls have been set by the Super
Kamiokande \cite{Takenaga2007} experiment and
MACRO \cite{ambrosio2002}.  The limits from Super Kamiokande are only
valid for relatively small values of the cross set ion (and therefor
mass and charge) due to the requirement that a Q-ball not interact in
the veto region of the detector.  Current neutrino detectors such as
IceCube and future gamma-ray and cosmic-ray detectors such as HAWC and
JEM-EUSO (note that JEM-EUSO has not published a sensitivity to Q-balls at the time of this report)
should be able to significantly improve upon the current
limits and extend the search to more theoretically interesting
parameter space of large cross sections.  Figure \ref{fig:qball-limit}
shows the current limits with the expected sensitivity of HAWC,
IceCube.  The diagonal lines are theoretical
expectations of for SUSY breaking scales of 1 and 10 TeV.  In any of
these detectors one should be able to determine the direction of a
Q-ball.  In this scenario, these instruments serve as directional
direct dark matter detectors.

%% file: CF6/strangelet.tex
\subsubsection{Strangelets}

Quark nuggets, nuclearites, and strangelets are different names for
lumps of a hypothetical phase of absolutely stable quark matter, named
{\it strange quark matter} because it is made of about one-third up,
down, and strange quarks ~\cite{Bodmer:1971we,Witten:1984rs}. Whether
strange quark matter (SQM) is absolutely stable is a question yet to
be decided by experiment or astrophysical observation (see
\cite{Alcock:1988re,Madsen:1998uh,Weber:2004kj,Madsen:2006qu} for
reviews). If stable, strange quark matter objects may exist with
baryon numbers ranging from ordinary nuclei to about $2\times 10^{57}$
corresponding to gravitational instability of strange stars
\cite{Alcock:1986hz,Haensel:1986qb}.

Small SQM lumps with baryon
number $A < 10^7$ are called strangelets \cite{Farhi:1984qu}.
Strangelets have an electron cloud neutralizing the slightly positive
quark charge. They are unlikely to survive the early Universe, but may
form at collisions of strange star binaries or may be accelerated off
the surface of pulsars. Relics from the cosmological quark-hadron
phase transition are usually called quark nuggets, while nuclearites
are nuggets that hit the Earth and may leave detectable signatures as
unusual meteor-events, earth-quakes, etched tracks in old mica, in
meteorites and in cosmic-ray detectors \cite{DeRujula:1984ig}.

Current searches for SQM states have excluded quark nuggets as dark
matter candidates with $3 \times 10^7 < A < 5 \times10^{25}$, but a
lower flux of relics or strange star collision debris has not been
ruled out. Several experiments have searched for strangelets in cosmic
rays with some interesting events not claimed as discoveries but
interpreted as flux limits reviewed here
\cite{Sandweiss:2004bu,Finch:2006pq}.  More recently the Lunar Soil
Strangelet Search (LSSS) has reported new strangelet limits
\cite{Han:2009sj} in a sample of 15 grams of lunar soil from Apollo
11.  The Alpha Magnetic Spectrometer (AMS) is currently searching for
stragelets from the International Space Station \cite{Kounine:2012zz}.
AMS is uniquely suited to discover extreme rigidity strangelets and
should be able to probe a wide mass range. AMS recently released their
first results \cite{Aguilar:2013qda} and have presented accurate
measurements of cosmic ray electrons, positrons, protons, and nuclei
at the International Cosmic Ray Conference 2013. No results from the
strangelet searches has been announced to date.

%% file: CF6/auger-upgrade.tex
\paragraph{Auger}

The Pierre Auger collaboration is currently planning an upgrade of the
existing $3000 km^2$ hybrid observatory located near Malarg\"ue,
Argentina.  The current observatory is comprised of an array of 1600
12 ton water Cherenkov surface detectors, most of which are spaced on
a 1.5 km hexagonal grid, together with air fluorescence detectors
distributed at four sites overlooking the surface detector array. The
objectives of the planned upgrade are to elucidate the origin of the
flux suppression at the highest energies, measure the composition of
UHECRs up to highest energies with sufficient resolution to detect a
10\% proton component, provide composition-tagging to facilitate
anisotropy studies, and study hadronic interactions at center-of-mass
energies an order-of-magnitude greater than at the LHC.

To address these objectives, the Auger upgrade focuses on enhancing
the ability of the detector to better separately determine the lateral
distribution function of the muon and electromagnetic (EM) components
of showers, and to extend these measurements to smaller core distances
(and large shower particle fluxes) than is possible with the existing
system. The upgrades also would improve measurement not only of the
depth of shower maximum for the EM component of the shower (as is done
at present using the fluorescence detectors in hybrid events) but also
the depth at which muon production reaches its maximum, using the
surface detector alone.

An upgrade of the electronics of the surface detectors is the first
step in separating the muon and EM components of the showers on an
event by event basis [3]. It includes faster timing of surface
detector signals, improving significantly the ability to distinguish
close-in-time muon pulses across the entire array. The updated
electronics will also facilitate the addition of dedicated muon
detectors over all or part of the surface detector array to further
improve the muon/EM separation and reduce model-dependent systematic
errors. This gives rise to a `boot-strap' approach, where the
model-independent, direct muon determination abilities of the upgraded
detector are used to validate and refine the more indirect detection
methods and analyses.

Distinct strategic options appear in how to use muon identification
and/or separation of the muonic and EM components of the signal, to
test and improve the hadronic physics modeling and to assign a
composition-probability to each event. The most direct method is to
simply separate the muonic and EM signals in each tank, and use timing
and geometry to infer the location of peak muon production point in
the development of the air shower[4]. A second approach is to fit the
total signal in every tank, including some elements of the FADC
timing, to a superposition of templates of EM and muonic
components. This gives a more accurate energy and angular
reconstruction than the traditional method, and at the same time gives
$X_{max}$ with remarkable accuracy , about $30\ g\ cm^{-2}$.

The additional muon identification technologies under study and
prototyping include (1) segmentation of the interiors of surface
detectors, to separate penetrating muons from the lower-energy
electromagnetic component, and (2) placement of external particle
detectors (such as RPCs or scintillators) with the existing Auger
detectors.

In addition, it is planned to extend
fluorescence light measurements into twilight by running with reduced
gain, thereby increasing the duty cycle for the highest energy showers
by up to a factor of two. This does not involve a hardware upgrade but
rather is an operational change in the existing detector.

It is planned to operate the upgraded Pierre Auger Observatory from
2015 to 2023, which would approximately double the data set which will
have been collected prior to implementation of the upgrade.

%% file: CF6/telescope-array.tex
\paragraph{Telescope Array}
\label{sec:ta_upgrades}

The Telescope Array (TA) collaboration has several upgrades in
progress and planned for the near future. The TA Low Energy (TALE)
extension will extend the observatory's reach in energy spectrum and
composition studies into the $10^{16}$~eV decade, enabling TA to probe
the ``second knee'' region and the galactic-to-extragalactic
transition. An additional decade downward in energy will be achieved
via the Non-Imaging CHErenkov (NICHE) array, bridging the gap between
Telescope Array and experiments operating in the knee regime. At the
highest energies, the TA$\times$4 (``TA times four'') detector will
greatly enhance Northern Hemisphere statistics, of particular
importance in arrival direction anisotropy and composition studies.

\subparagraph{\it TALE} While TA has been able to extend analysis down to about
$10^{18}$~eV, this is insufficient to fully observe the
galactic-to-extragalactic transition. In addition, it is optimal to
observe cosmic rays from LHC energies through the second knee and up
to the GZK cutoff with one well cross-calibrated detector. TALE, the
low energy extension to the Telescope Array, is designed to lower the
energy threshold to about $10^{16.5}$~eV. To do this, an additional
ten fluorescence telescopes were installed,
viewing up to 57 degrees in elevation angle. Installation of a new
graded array of about 100 scintillator detectors is currently in
progress. This extension will enable the Telescope Array to measure
the energy and composition of cosmic rays to much lower energies while
cross calibrated with the detectors of the main Telescope Array. By
pushing the energy threshold down to $10^{16.5}$~eV, the TA
collaboration hopes to sort out the galactic and extragalactic
contributions to the cosmic ray flux.

\subparagraph{\it NICHE} Co-sited with TA/TALE, the Non-Imaging CHErenkov Array
(NICHE) will measure the flux and nuclear composition of cosmic rays
from below $10^{16}$~eV to $10^{18}$~eV in its initial
deployment. Furthermore, the low-energy threshold can be significantly
decreased below the cosmic ray knee via counter redeployment or by
including additional counters. NICHE uses easily deployable detectors
--- consisting of a single phototube and Winston cone --- to measure
the amplitude and time-spread of the air-shower Cherenkov signal to
achieve an event-by-event measurement of $X_{max}$ and energy, each
with excellent resolution. NICHE will have sufficient area and angular
acceptance to have significant overlap with the TA/TALE detectors to
allow for energy cross-calibration. Simulated NICHE performance has
shown that the array has the ability to distinguish between several
different composition models as well as measure the end of Galactic
cosmic ray spectrum.

\subparagraph{\it TA$\times$4} The Telescope Array (TA) collaboration is proposing
to build TA$\times$4. This is a project to expand the TA surface
detector by a factor of four, bringing the total instrumented area to
3000~km$^2$.  The plan is to build 500 more scintillation counters and
deploy them in an array of 2.08~km spacing.  This array, plus the
existing TA SD would reach the design size.  There is plenty of room
at the TA site to expand on the northern, western, and southern sides
of the TA SD.  The new array would need a fluorescence detector
overlooking it to set the energy scale, so the TA$\times$4 plan
includes a fluorescence detector of 10~telescopes.  These will be
reconditioned HiRes telescopes.

The aim of the design is to collect data for anisotropy studies at the
highest energies.  An anisotropy signal due to the local large scale
structure of the universe (the local 250 Mpc) really should be present
at energies larger than about~57 EeV, where the extragalactic and
galactic magnetic fields have an effect on cosmic rays’ trajectories
smaller than the size of the local large scale structures.  Telescope
Array has already seen hints of such structure. If successful, the TA$\times$4
project will take 3 years to acquire the funds, build and deploy the
detectors. Including the current 5~TA-years of data, 3~years of
operation of TA$\times$4 would yield 20~TA-years of data. These data
will be sufficient to determine unambiguously whether an LSS
anisotropy exists.  If the LSS signal really comes from the TA hot
spot, 3~years of TA$\times$4 data will clarify this signal and yield a
$5\sigma$ observation.

%% file: CF6/JEM-EUSO.tex
\paragraph{JEM-EUSO}

JEM-EUSO will be the first space observatory for the study of extreme
energy cosmic rays with energies of $\sim 10^{20}$ eV. The Extreme
Universe Space Observatory (EUSO) to be accommodated on the Japanese
Experiment Module (JEM) of the International Space Station (ISS) will
look down towards the Earth and use the atmosphere as a giant detector
\cite{Adams:2013vea,TheJEM-EUSO:2013vea}. The 2.5 m ultraviolet (UV)
telescope with a 60$^o$ field of view (FOV) will observe the
fluorescence signal produced by the extensive air-showers (EAS)
generated by extremely energetic cosmic rays (EECRs) that enter the
EarthÕs atmosphere.

The main objective of JEM-EUSO is to identify the sources of the
highest energy cosmic rays and thus begin particle astronomy. JEM-EUSO
will significantly increase the worldwide data collection of particles
at extreme energies providing clear anisotropy signals for the
identification of the first sources of extragalactic cosmic rays and
the measurement of the energy spectrum beyond the
Greisen-Zatsepin-Kuzmin (GZK) feature. Identifying the sources will
solve a longstanding mystery and further the study of particle
interactions with center of mass energies beyond 100 TeV.

JEM-EUSO is also sensitive to very low fluxes of extremely high-energy
neutrinos that may be produced if cosmic accelerators reach higher
energies than those observed thus far. The observation of extremely
energetic neutrinos would make possible studies of neutrino
interactions with center of mass above 100 TeV.

JEM-EUSO will also contribute to the investigation of phenomena
intrinsic to the EarthÕs atmosphere or induced by the flux of
meteoroids and strangelets (or nucleorites) incoming from space.

A worldwide collaborating of 75 research groups from 13 countries is
designing JEM-EUSO to operate for more than 3 years onboard the ISS
which orbits around the Earth every 91 minutes at an altitude of about
400 km. JEM-EUSO will image light from the isotropic nitrogen
fluorescence excited by the EAS, and the forward-beamed Cherenkov
radiation reflected from the EarthÕs surface or cloud tops. The
highly-pixelized high-speed JEM-EUSO camera will capture the time
development of an EAS to determine the energy and arrival direction of
the EECR.  The cameraÕs focal plane is covered by MAPMTs with $3
\times 10^5$ pixels, each less than 3 mm, giving a 0.07$^o$ resolution
per pixel; a pixel covers about 0.5 km on the surface of the
Earth. The characteristics of the EAS can be used to determine the
original direction, energy, and nature of the EECR. JEM-EUSO will be
able to discriminate between hadron, gamma ray, and neutrino initiated
showers.

%% file: CF6/CR-radio.tex
\paragraph{Radio Detection of Cosmic Rays}  

Detection of cosmic ray air showers using radio techniques has
undergone a resurgnece of interest in recent years.  Detectors have
been operated in coincidence with air shower arrays at Auger
\cite{Berat:2013vva,Kleifges:2013fwa}, Telescope Array
\cite{Ogio:2013oza}, and Kascade-Grande
\cite{Smida:2013sza,Apel:2013kda}.  Cosmic ray radio emission has also
been detected by the ANITA balloon-borne
interferometer\cite{Hoover:2010qt}.  The dominant mechanisms include
geosynchrotron emission and Cherenkov emission, and result in
polarized emission beamed near the shower axis.  The intensity and
lateral distribution of the radio signal can be used to infer shower
energy and depth of maximum\cite{Apel:2013kda}.  In addition detectors
being planned in conjunction with air shower arrays, there are
prospects for radio detection using large interferometric array such
as LOFAR\cite{vanHaarlem:2013dsa,Nelles:2013fua} and SKA, as well as
from space\cite{Romero-Wolf:2013etm}.

%% file: CF6/CR-radar.tex
\paragraph{Radar Detection of Cosmic Ray Showers}

Radar is a another candidate remote sensing technique with a potential
to achieve a 100\% duty cycle. The Telescope Array RAdar (TARA)
project in Utah is designed to test the idea that ionization
produced by extensive air showers should scatter RF radiation.

TARA employs two analog television transmitters, with a combined
output power of 40~kW, broadcasting a 54.1~MHz (low-VHF) signal over
the Telescope Array surface detector. The signal is enhanced by use of
a high gain (over 20~dBi) phased array of Yagi antennas which boosts
the equivalent isotropic radiated power to over 8~MW.

Due to the high velocity of the air shower ionization front, the RF
scattered off of an extensive air shower will be characterized by a
Doppler shift of several tens of Megahertz. To detect such signals,
TARA employs a 250~MS/s receiver along with an onboard FPGA to allow
smart triggering at a level well below that of galactic sky noise.

TARA commissioning is currently underway, with preliminary results
anticipated in Fall of 2013.

%% file: CF6/gammaexp.tex

\subsubsection{Future Gamma-Ray Experiments}
\label{sec:CF6-gammaexp}

At energies above 100 GeV the flux of gamma rays from astrophysical
objects is sufficiently small that detection using a space-based
instrument becomes prohibitively expensive.  Ground-based VHE
gamma-ray instruments are of two types: Imaging Atmospheric Cherenkov
Telescopes (IACTs), that use large mirrors to image the Cherenkov
light generated in the atmosphere by extensive air showers and
Extensive Air Shower (EAS) arrays, that directly detect particles that
reach the ground - predominantly gamma rays, electrons and positrons.
The characteristics of the two types of instruments complement each
other and together they provide excellent coverage of a broad array of
astrophysical objects.  IACTs have significantly better instantaneous
sensitivity, better energy resolution, and better angular resolution.
However, they can only view a small portion of the sky at any one time
and they can only operate on clear moonless nights (though progress
has been made in extending operations into the lunar cycle).  (Current
instruments have a field-of-view of $\sim$5milli-sr and the planned
CTA will have a field-of-view of $\sim$15 milli-sr.)  In contrast, EAS
arrays can operate continuously and are sensitive to air showers from
the entire overhead sky.  Because the energy threshold of EAS arrays
depends upon the atmospheric depth and therefore the zenith angle of
the primary gamma ray, field-of-view is typically considered to be 2
sr.  In IACTs the rejection of the cosmic-ray background is
accomplished through a combination of image analysis and angular
resolution.  In EAS arrays background rejection is accomplished
through a combination of muon detection and angular resolution.

Given the above characteristics IACTs are to:
\begin{itemize}
\item perform detailed morphological studies (spectrally resolved) of
  Galactic sources for multi wavelength studies - required to
  understand acceleration mechanisms
\item extend ground-based measurements to relatively low energies
  (below 50 GeV) - required to probe the pulsar mechanism, provide
  overlap with space-based instruments, and for sensitivity to lower
  has WIMPs
\item observe the fastest transients from known sources - required to
  understand the acceleration mechanism and environment in active
  galaxies and provide the best limits on violation of Lorentz
  invariance
\item detect faint sources and thereby provide the best indirect
  limits on dark matter annihilation
\item resolve energy spectra of astrophysical sources - required for
  dark matter detection, understanding acceleration processes, and
  searching for evidence of axion-like particles
\end{itemize}

Similarly EAS arrays:
\begin{itemize}
\item monitor the sky and alert the more sensitive IACTs and other
  instruments operating at different wavelengths (and particle type)
  for followup observations of transient phenomena - critical to
  understanding complex astrophysical environments).
\item detect prompt VHE emission from gamma-ray bursts
\item perform unbiased sky surveys to discover new sources and
  phenomena (including undetected high M/L dark matter sources)
\item study the highest energy gamma rays, where sensitivity is
  typically flux limited
\item study large sources such as galaxy clusters (for dark matter
  annihilation) and Galactic and extragalactic diffuse emission
\end{itemize}

Given the complexity and diversity of sources of VHE gamma rays it is
important to have both types of instruments available as well as
instrument operating at lower wavelengths (100 MeV gamma rays, x-rays,
radio, and optical) to glean the full benefit from any of the
instruments.  Single observations at a single wavelength are rarely
(if ever) sufficient to properly understand astrophysical sources - a
requirement if one is to extract fundamental physics from VHE
gamma-ray observations.  Therefore it is important to operate Fermi,
VERITAS, and HAWC simultaneously for a period of several years.
Results from such a period will inform us of the value of these
observations and allow one to make physically motivated decisions
about further operations.

\paragraph{VERITAS}
VERITAS (the Very Energetic Radiation Imaging Telescope Array System)
is a ground-based gamma-ray instrument operating at the Fred Lawrence
Whipple Observatory in southern Arizona, USA. It comprises an array of
four 12m optical reflectors, which exploit the imaging atmospheric
Cherenkov technique to measure emission from astrophysical sources in
the $\sim100$ GeV to $\sim30$ TeV energy range. The array has been
operating smoothly since 2007, recording around 1000 hours of data per
year. The angular and energy resolution are energy dependent: at 1 TeV
they are $\sim0.1^{\circ}$ (68\% containment radius) and $\sim15\%$,
respectively. The sensitivity of the array is sufficient to detect a
source with 1\% of the steady Crab Nebula flux in less than 25 hours,
while the Crab itself is detected in a matter of seconds. VERITAS has
recently completed a series of upgrades, which included relocating the
original prototype telescope, installing a new trigger system, and
upgrading the telescope cameras with more sensitive photomultiplier
tubes. These upgrades combined have halved the time required to detect
a typical gamma-ray source.

VERITAS has now detected 46 sources - around one-third of the known
TeV catalog - and over half of these detections were new
discoveries. The catalog comprises active galaxies (blazars and radio
galaxies), a starburst galaxy, a pulsar, many pulsar wind nebulae,
binary systems, supernova remnants, and sources whose nature remains
unknown. Core science goals include the study of cosmic ray particle
acceleration, both within our Galaxy (e.g. in supernova remnants) and
externally (e.g.  in AGN jets), searching for the gamma-ray signature
of dark matter annihilation and primordial black hole evaporation, and
constraining fundamental physical effects such as Lorentz invariance
violation. The attenuation and cascading of TeV gamma-ray photons from
distant sources can also be used to measure or constrain the
extragalactic infra-red background light, and the extragalactic
magnetic field strength. All of these studies are greatly enhanced by
contemporaneous overlap with complementary gamma-ray instruments
including the Fermi gamma-ray space telescope and HAWC.

\paragraph{Cherenkov Telescope Array}

The Cherenkov Telescope Array (CTA) is a concept for a ground-based
observatory \cite{cta2011} for very high-energy (VHE, 30 GeV - 300
TeV) $\gamma$ rays.  It will use imaging atmospheric Cherenkov
telescopes (IACTs) deployed over $\ge$10 km$^2$ to detect flashes of
Cherenkov light from air showers initiated by $\gamma$ rays, a
technique pioneered at Whipple Observatory in the US.  Current IACTs
have cameras consisting of $\sim$ thousand fast photomultiplier tubes,
and their effective collecting areas typically reach $\sim$~0.1
km$^2$.  The current generation, namely H.E.S.S., MAGIC, and VERITAS,
have up to 5 IACTs with separations of $\sim$100~m to record multiple
views of each shower. CTA will be an array of $\sim$~one hundred IACTs
to increase the collection area and the number of recorded images of
each $\gamma$-ray shower.

The current concept for CTA consists of subarrays with telescopes of 3
different sizes. The CTA-US consortium is working towards contributing
mid-sized telescopes (MSTs, 9 - 12 m) to CTA (e.g., 36 MSTs were
recommended by Astro2010 \cite{blandford2010} and PASAG
\cite{ritz2009}) to optimize the performance of CTA in the 80~GeV -
5~TeV regime.  The U.S. groups are currently designing and building a
prototype telescope to test the feasibility of an innovative,
Schwarzschild-Couder telescope (SCT) design \cite{schwarzschild1905,
  vassiliev2007, vassiliev2008} that will allow much smaller, less
expensive cameras, yet providing $\sim$ 10$^4$ pixels, better angular
resolution and a much larger field of view than conventional
Davies-Cotton designs.  An important aspect of the prototype program
is the design and testing of a new optical system, mirror technologies
and camera electronics for the necessary advances in performance,
reliability and lowered cost.

\begin{figure}[ht]  
	\centerline{\includegraphics[width=5.5in, height=3.5in, angle=0.0]{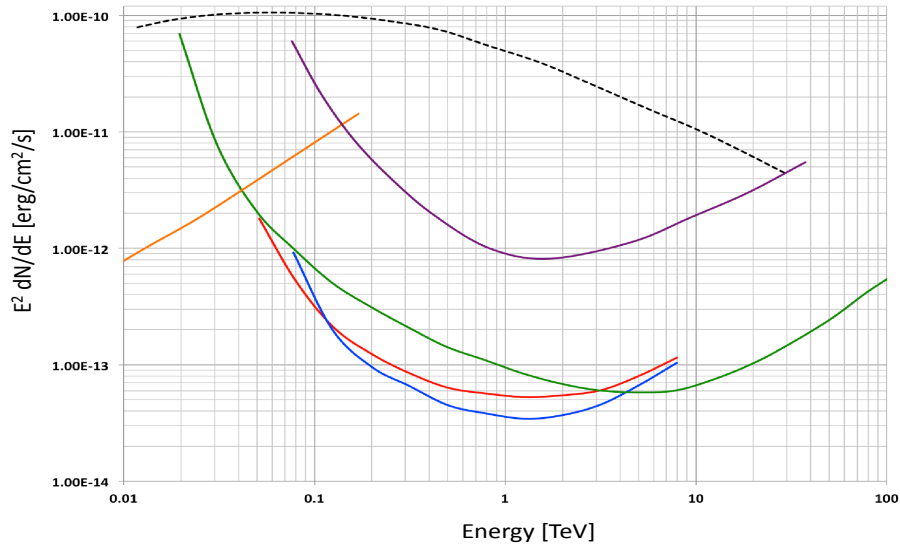}}
	\caption{\footnotesize \sf\textsl - The dashed line shows the
          differential energy spectrum of the Crab Nebula.  The solid
          lines depict the sensitivity for a 50~hour exposure ($\ge
          5\sigma$ detection and $\ge$~10 photons per energy bin) for
          various configurations of CTA and the current generation
          H.E.S.S. and VERITAS (solid purple) observatories. The CTA
          baseline array is shown in green using the best performance
          of any of the arrays considered in \cite{bernloehr2013}.
          The redline shows the addition of the US contribution to a
          total of 61 mid-sized telescopes (e.g., 25 baseline and 36
          CTA-US telescopes) based on the DC design, whereas the blue
          line show the addition of a US contribution with a total of
          61 mid-sized telescopes based on the SC design.  The
          sensitivity for Fermi is shown (solid orange) for an
          exposure time of 10 years.
	\label{fig:CTASensitivity} }   
\end{figure}

This work is a collaboration between national laboratories,
universities and industry and consists of $\sim$~20 research groups in
the US.  The US groups have received ~5M\$ through an NSF-MRI program
to construct a prototype telescope between 2012-2015 with the goal to
get a realistic estimate of construction costs and performance.

Overall, CTA will (a) provide an order of magnitude better sensitivity
for deep observations ($\sim 10^{-3}$~Crab nebula flux); (b) have a
much greater detection area ($\sim$~1 km$^2$), and hence detection
rates, for transient phenomena; (c) improve the angular resolution
(0.02$^{\circ}$ at 1 TeV) to resolve cosmic accelerators; (d) provide
uniform energy coverage from $\sim$~30 GeV to beyond 100 TeV photon
energy; and (e) enhance the sky survey capability, monitoring
capability, and flexibility of operation relative to current IACTs.
These improvements will provide a dramatic step in exploring
non-thermal processes in our Universe.

Figure \ref{fig:CTASensitivity} shows simulation results of the flux
sensitivity for several possible telescope configurations: the
baseline array \cite{bernloehr2013} (solid green line) using up to 25
mid-sized telescopes (MSTs) and the large-sized and small-sized
telescopes to cover the low energies (E $\le$~0.1~TeV) and high
energies (E $\ge$~10~TeV).  Furthermore we show the effect when adding
the US contribution of mid-sized telescopes (recommended by Astro2010
and PASAG), amounting to either 61 DC-MSTs or 61 SC-MSTs.

\begin{figure}[ht]
	\centerline{\includegraphics[width=3.5in, height=5.0in, angle=90.0]{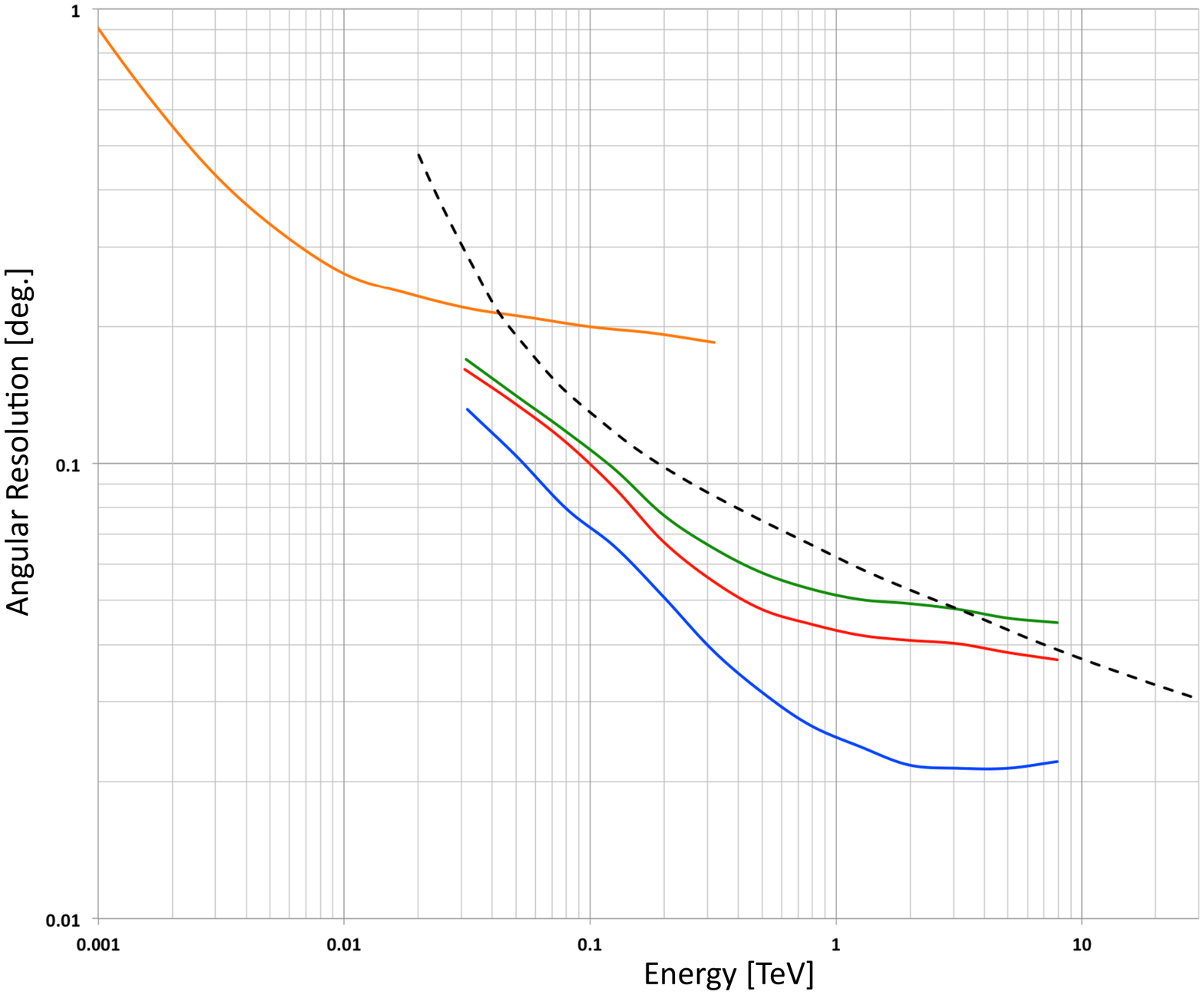}}
	\caption{\footnotesize \sf\textsl - The angular resolution is
          shown the CTA baseline array (solid green), the addition of
          the US contribution of mid-sized telescopes, for 61 DC-MSTs
          (solid red) and 61 SC-MSTs (solid blue).  For comparison we
          also show the angular resolution of Fermi (solid orange).
	\label{fig:CTA-AngRes} }   
\end{figure}

The addition of the US telescopes would bring the sensitivity to the
$\sim 10^{-3}$~Crab level, and improve the sensitivity up to a factor
of 3 in the 80~GeV - 5 TeV regime.  This large improvement is due to
increasing the MST array to its optimal size, where a much larger
fraction of events fall within the array (contained events) for
excellent event reconstruction, and the better resolution SC telescope
compared to the DC design.

Figure \ref{fig:CTA-AngRes} shows the results of a simulation of the
angular resolution for several telescope configurations for CTA as a
function of energy. The black dashed line shows the CTA requirement
for angular resolution \cite{ctarequirement}, the green line
corresponds to the baseline array.  The blue line and the red line
show the angular resolution with and without the US contribution of more than doubling the number of mid-sized 
telescopes, the 61 SC-MSTs or 61 DC-MSTs, respectively.  For
comparison we also show the angular resolution of Fermi (orange line).
It should be emphasized that the use of the SC design provides a
substantial improvement in angular resolution, which not only
translates into a better sensitivity for point sources, but also
potentially provides new physics capabilities through high-resolution
imaging in the TeV regime.

\paragraph{The High Altitude Water Cherenkov Experiment}
The High Altitude Water Cherenkov Experiment (HAWC) is a wide
field-of-view ($>$2 sr), high duty-cycle ($>95\%$) TeV gamma-ray
experiment currently under construction at the vulcan Sierra Negra in
Mexico.  HAWC will be composed of 300 large water Cherenkov detectors
(WCDs) (7m diameter by 4.3 meters high), instrumented with 4
photomultiplier tubes (PMTs).  Three of the PMTs are 8" Hamamatsu
R5912 (reused from the Milagro experiment) and the 4$^{th}$ central
PMT is a high quantum efficiency 10" R7081-MOD Hamamatsu PMT.  The
WCDs will be placed in a close-packed array covering a total area of
approximately 20,000 m$^2$.  The array will begin operations with 100
WCDs in August of 2013 and the full array should begin operations
approximately one year later.  HAWC builds upon the success of the
Milagro experiment, which demonstrated the advantages of using water
Cherenkov technology for wide-field ground-based gamma-ray instruments
(dense sampling of the air shower particles on the ground and
sensitivity to the gamma rays in an air shower) that lead to a
dramatically lower energy threshold over previous generation of
instruments based upon a sparse array of plastic scintillators.  In
five years of operation HAWC will survey 8 sr of the sky with a
sensitivity roughly 15 times greater than that of Milagro and well
matched to the sensitivity of current IACTs.

The sensitivity of HAWC to point sources of TeV gamma rays will be
roughly 15 times that of Milagro, enabling HAWC to detect the Crab
Nebula in a single transit (compared to 3-4 months for Milagro).  As
an all-sky instrument with a low energy threshold below 100 GeV, HAWC
is well suited to detect transient phenomena in the VHE sky.  Flares
from active galaxies and gamma-ray bursts are prime scientific targets
for HAWC \cite{abeysekara2012}.  Extrapolations from current
detections of GRBs by the Fermi LAT indicate that HAWC should detect
about 1.6 GRBs each year \cite{taboada2013}.  The detection of a GRB
at $>$100 GeV would enable HAWC to probe for violations of Lorentz
Invariance with a sensitivity about a factor of two beyond current
limits.  Flares (or lack of flares) from distant AGN will directly
test the UHECR origin of the highest energy gamma rays from these
objects, which is critical to understanding the sources of UHECRs and
the role of axion-like particles in the propagation of VHE gamma rays.
The final state emission from a primordial black hole will have a
signature that is similar too - but distinct from - a gamma-ray burst
and HAWC will be sensitive to such emission over a volume roughly 2
orders of magnitude larger than has been probed to date.  HAWC's
sensitivity to the annihilation of dark matter particles peaks in the
dark matter mass range above 10 TeV and the large field-of-view gives HAWC unique
sensitivity to baryon-poor dwarf galaxies for high-mass dark matter.

\paragraph{LHAASO}
The LHAASO (Large High Altitude Air Shower Observatory) is an ambitious project based upon a combination of water Cherenkov technology, scintillation detectors, and air Cherenkov technology.  LHAASO will consist of a $\sim$90,000m$^2$ water Cherenkov detector surrounded by 5100 scintillation detectors distributed over an area of $\sim$1km$^s$ with 43,000 m$^2$ of buried muon detectors.  In addition 24 air fluorescence/Cherenkov telescopes will be located onsite.  At an altitude of $\sim$4300m, it is expected that LHAASO will have somewhat better sensitivity than HAWC at low energies ($<$10 TeV), with significantly improved sensitivity at higher energies.  This project recently received approval from the Chinese government and the completion of construction is expected in 2018.


\paragraph{A Future Wide-Field High-Duty Cycle Gamma-Ray Experiment}
HAWC was designed and built based on the results from the Milagro experiment.  Similarly, the design of a future wide-field high duty-cycle experiment will be based upon the results from HAWC (or LHAASO).  There are two distinct paths for a future instrument: significantly higher sensitivity to higher energy gamma rays, in excess of 100 TeV or significantly reducing the useful energy threshold.  If the HAWC data shows that exciting physics is to be found at the highest energies (cosmic-ray origins, Galactic gamma-ray sources), then a plan to increase the collecting area at the highest energies would be recommended.  Such an upgrade could be performed at the existing HAWC site or at a new location, perhaps in the Southern hemisphere to provide an alert system for the CTA South.  On the other hand, if extragalactic phenomena, especially transient events such as flares from active galaxies to gamma-ray bursts, yield a rich source of information on particle acceleration, ultra-high-energy cosmic rays, and tests of fundamental physics, a detector with a significantly lower energy threshold would be recommended.  Such an instrument would require the highest altitude site attainable, and thus would naturally be placed in the Southern hemisphere.  Within the Chajnantor plateau in Chili it seems feasible to site such an instrument at $\sim$6km above sea level. 

%% file: CF6/icecube.tex
\paragraph{IceCube and KM3NeT}

IceCube is a 1 km$^3$ neutrino observatory located at the South Pole
\cite{Halzen:2010yj}.  Completed in December, 2010, it instruments 1
km$^3$ of Antarctica ice with 5,160 optical sensors, mounted at depths
between 1450 and 2450 m, on 86 vertical strings which are emplaced in
holes drilled in the icecap.  These sensors observe the Cherenkov
radiation from the charged particles which are produced when
high-energy neutrinos interact in the Antarctic ice. 78 of the strings
are arranged on a 125 m triangular grid; this array has an energy
threshold of about 100 GeV.  The remaining 8 strings form a denser
subarray called ``DeepCore" \cite{Collaboration:2011ym}; these strings
have smaller spacings, with most of DOMs on the bottom 350 m of the
strings. DeepCore has an energy threshold of about 10 GeV.

A surface array, called IceTop, completes the installation
\cite{IceCube:2012nn}.  It comprises 162 ice filled tanks.  The array
is sensitive to cosmic-ray air showers with energy above about 100
TeV.  One key feature of IceTop is its altitude; at 2735 m above sea
level, so IceTop is relatively near shower maximum for the showers of
greatest interest (above 1 PeV); this reduces its sensitivity to many
systematic errors, such as hadronic interaction models.

IceCube observes about 200 neutrino interactions per day, mostly
produced in cosmic-ray air showers. It has measured the $\nu_\mu$
spectrum from energies of 100 GeV up to 1 PeV, the atmospheric $\nu_e$
spectrum from energies of 80 GeV up to 6 TeV, and has set limits on
$\nu_\tau$ \cite{Abbasi:2012cu}.  Using DeepCore, it has observed
atmospheric neutrino oscillations, studying neutrinos with energies in
the 10-60 GeV region \cite{Aartsen:2013jza}.

IceCube has searched for astrophysical neutrinos in many channels,
including searches for point sources \cite{Aartsen:2013uuv}, episodic
sources, GRBs \cite{Abbasi:2012zw} and diffuse searches.  The point
source searches have not observed any excesses over atmospheric
backgrounds, but the diffuse searches have seen a clear excess of
expectations, most notably "Bert" and "Ernie," two neutrino events
with energies above 1 PeV \cite{Aartsen:2013bka}.  A systematic study
found 28 contained events, with energies above 60 TeV, over an
expected background of about 12 events; this is roughly a $4\sigma$
excess \cite{IceCubeICRC}.

IceCube has also studied cosmic-ray air showers, including
measurements of the energy spectrum, composition (combining IceTop air
shower data with buried measurements of TeV muon fluxes) and
anisotropy \cite{IceCubeICRC2}.  Surprisingly, the anisotropy persists up to
energies of at least 400 TeV; this challenges our models of cosmic-ray
production and propagation \cite{Abbasi:2011zka}.  It also studies
high transverse momentum muons produced in air showers, establishing
the connection between cosmic-ray physics and perturbative quantum
chromodynamics \cite{Abbasi:2012kza}.

IceCube also searches for a variety of beyond-standard model
phenomena: neutrinos from WIMP annihilation in the Sun, the Earth, the
galactic halo or nearby dwarf galaxies, searches for magnetic
monopoles and upward-going particle pairs; the latter is expected many
variants of supersymmetry with a high mass scale.

A consortium of European institutions are proposing to build the
KM3NeT detector \cite{Kooijman:2013qla}.  This 5-6 km$^3$ detector
would be located in the Mediterranean sea, where it would have a good
(neutrino) view of the galactic center.  Seawater has a somewhat
longer optical scattering length than Antarctic ice, so KM3NeT should
also have somewhat better angular resolution than IceCube.

%% file: CF6/pingu.tex
\paragraph{PINGU and other high-density detectors}

PINGU(Precision IceCube Next Generation Upgrade) is a proposed
high-density infill array within IceCube and DeepCore
\cite{IceCube:2013aaa}.  It will consist of 20 to 40 additional
strings, with at least 60 optical modules per string, to observe
neutrinos with energies down to a few GeV.  The optical modules would
be similar to those used in IceCube.  Its main physics goal is to
determine whether the neutrino mass hierarchy is 'normal' or
'inverted.'  It is sensitive to the hierarchy because low energy
electron neutrinos passing through dense matter (i.e. the Earths core
and mantle) can resonantly oscillate into other flavors; the energy
dependence of this resonant oscillation depends on the hierarchy.

ORCA (Oscillation Research with Cosmics in the Abyss) \cite{ORCA}is a
proposed high-density phototube array in the Meditranean.  It would
study a similar set of physics topics as PINGU.

Looking further ahead, MICA (Multi-Megaton Ice Cherenkov Array) would
be an array with an even denser array of photosensors
\cite{Deyoung:2013rla}.  It will instrument a large volume (several
megatons) with enough photosensor area to be search for proton decay
and observe supernova neutrinos from other galaxies.  The supernova
search will benefit from both the large volume and the high photon
sensor density to allow nearly background-free searches for neutrinos
from supernovas in moderately nearby galaxies; the goal is a detector
big enough to will collect a statistically meaningful number of
supernovae.

%% file: CF6/nu-askaryan.tex
\paragraph{Radio Cherenkov Experiments}

Radio Cherenkov experiments exploit the Askaryan emission produced by
the excess negavie chage which develops in shoers ocurring in dense
media. Active and proposed radio Cherenkov experiments can been
classified as balloon-borne and {\it in situ}.  Balloon experiments
have a higher threshold, but can view an enormous volume of ice from
their high altitudes.  Radio arrays {\it in situ} expect be able to
reconstruct the neutrino locations and directions better than balloon
experiments due to their close proximity to the interactions.  Balloon
experiments view the neutrino sky at low declinations while {\it in
  situ} arrays can view downward and Earth-grazing directions.

\subparagraph{\it ANITA}

The ANITA (ANtarctic Impulsive Transient Antenna) is a balloon-borne
radio Cherenkov experiment designed to search for radio impulsive
signals induced by UHE neutrino interactions from $\sim$37~km altitude
above the Antarctic ice.  ANITA has flown twice (ANITA~1 in the
2006-2007 season and ANITA~2 in 2008-2009) under NASA's long-duration
balloon program, and has a third flight approved and planned for the
2014-2015 season.  ANITA flies broadband (200~MHz-1200~MHz), dually
polarized antennas that view the typically $1.5$~million km$^2$ of ice
in view of the payload. The signature for a UHE neutrino interaction
in ANITA would be a set of impulsive signals that are not consistent
with being from any known base or other human activity, or any other
event.  From the non-observation of a neutrino signal in its first two
flights, ANITA places the world's best constraints on the UHE neutrino
flux above $10^{19}$~eV.  ANITA~3 will fly 48 antennas (compared to 40
for ANITA~2) with improved response at the crucial low end of the
band, and will for the first time perform cursory interferometric
analysis at the trigger level, allowing for a $\sim20-30$\% reduction
of the threshold.

\subparagraph {\it ARA}
  
The Askaryan Radio Array (ARA) is an {\it in situ} array being
deployed near the South Pole.  The ARA collaboration aims to detect of
order 10 UHE neutrinos per year by instrumenting the 100's of~km$^3$
detection volume of ice with a 200~m deep array arranged in stations
of 16 antennas, horizontally and vertically polarized and bandwidths
approximately 150~MHz-800~MHz.  A prototype testbed station and the
first three ARA stations were deployed between the 2010-2011 and
2012-2013.  In the last season the ARA drill team successfully reached
the 200~m design depth below the snow-ice transition layer called the
firn.  The ARA collaboration is proposing to deploy another seven
stations in the 2013-2014 and 2014-2015 seasons, bring the detector to
10 stations (ARA10) on the path towards a ARA37 spanning 100~km$^2$ of
ice area.  ARA10 will have discovery potential for many data-driven
neutrino flux models while ARA37 will either begin to collect a sample
of neutrinos from those models, or reach even the more pessimistic
models where the cosmic ray composition at the highest energies is
mixed.

\subparagraph {\it ARIANNA}

ARIANNA (Antarctic Ross Ice-Shelf Antenna Neutrino Array) is a surface
array being deployed on the Ross Ice Shelf with a similar aim to reach
the detector volume necessary to measure a sample of order 100 UHE
neutrinos.  ARIANNA aims to detect the impulsive radio Cherenkov
signals from neutrino interactions both directly, and after
reflections from the highly reflective ice-sea water boundary below
the shelf.  Together the direct and reflected signals complete the
coverage of the upper sky.  Initial results using a broadband pulser
have verified the integrity of signals post-reflection.  ARIANNA is
on-track to complete a 7-station hexagonal detector unit in the
2013-2014 season and proposes to deploy a 960 station array.  Although
ARIANNA requires a large number of stations for the same detection
volume compared to a deep array like ARA, a surface array is in
principle simpler to deploy.
 
\subparagraph{\it EVA}
 
The ExaVolt Antenna (EVA) is a next-generation balloon experiment that
would turn the stadium-sized balloon itself into an enormous antenna,
which would make EVA the world's largest airborne telescope.  An
impulsive signal from the ice below would be incident on $\sim$10~m
high reflector region that would be affixed along the bulge of the
balloon, and focused onto the receiver array suspended on the inside
of the balloon.  EVA takes advantage of the Super Pressure Balloon
technology that is being developed by NASA, where the internal
pressure of the balloon is higher than the outside pressure and the
balloon holds its shape to $\sim$1\% along any dimension.  EVA expects
to use a 18.5~Mft$^3$ balloon for the full flight.  Microwave scale
models of EVA reflector sections built and tested at the University of
Hawaii have demonstrated a 23~dBi antenna gain and a focus region that
would scale to one to a few meters for the full balloon.  The EVA
collaboration was awarded a 3 year engineering grant and will carry
out a hang test of a 1/20 scale, 5~m diameter EVA prototype in the
Fall of 2013 or Spring of 2014 before proposing a flight of the full
EVA.  EVA would boast the best sensitivity to the neutrino spectrum at
the highest energies.

%% file: CF6/matter-assymmetry.tex
The Cosmological Asymmetry between matter and anti-matter provides
firm evidence for non-dark physics beyond the standard model. The
current net density of baryons implies an asymmetry during the hot
early universe between quarks and anti quarks, with about $10^8 +1$
quarks for every $10^8$ anti-quarks. This asymmetry must have arisen
after the end of inflation, due to some unknown mechanism called {\it
  baryogenesis}. In 1967 Andrei Sakharov proposed that baryogenesis
could arise from out of equilibrium new physics which violates baryon
number conservation, and the C and CP symmetries between matter and
antimatter\cite{Sakharov:1967dj}. At nonzero temperature baryon number
violation in the standard model proceeds via the baryon number
violating electroweak field configurations are known as {\it
  sphalerons}\cite{Klinkhamer:1984di}.  In 1985 Kuzmin, Rubakov and
Shaposhnikov pointed out that sphaleron processes are sufficiently
rapid at the high temperatures of the early universe to play a role in
baryogenesis \cite{Kuzmin:1985mm}.  They also proposed that a strongly
first order electroweak phase transition could provide the necessary
departure from thermal equilibrium. However it is now known that the
minimal standard model with a 125 GeV Higgs does not undergo a phase
transition\cite{Kajantie:1993ag,Jansen:1995yg,Kajantie:1996mn}, and
also does not have sufficient CP violation to produce an asymmetry of
order $10^{-8}$ \cite{Gavela:1994dt}. However many extensions of the
standard model do provide the necessary conditions for baryogenesis,
as well as exciting opportunities for a wide variety of experiments.
For recent reviews, see ref. \cite{Dolgov:1991fr,Dine:2003ax}. Here we
summarize the most well motivated possibilities.

\subsection {Leptogenesis}
The recent advent of the evidence of non-zero neutrino masses opens up
the possibility of {\it leptogenesis}~\cite{Fukugita:1986hr},
generation of a net lepton number. As sphaleron processes conserve
$B-L$, the difference between baryon number and lepton number, they
tend to convert part of the primordial lepton number asymmetry into a
baryon number asymmetry.  The successful implementation of
leptogenesis requires the existence of new CP violating phases in the
lepton sector. In this scenario, the baryon asymmetry is related to
the properties of the neutrinos. This subject is an example of the
synergy between physics at the Cosmic and the Intensity
Frontiers~\cite{Chen:2007fv}.  

\subsubsection{Standard Leptogenesis} 

Standard leptogenesis~\cite{Fukugita:1986hr} is implemented within the
seesaw mechanism for small neutrino masses in the presence of
right-handed neutrinos. The full seesaw Lagrangian contains the Yukawa
interactions for the neutrinos that in turn gives the Dirac mass terms
for the neutrinos, as well as the lepton number violating Majorana
mass terms, $M_{R}$, for the right-handed neutrinos.  At temperature
$T <~ M_{R}$, right-handed neutrinos, $N$, can generate a
primordial lepton number asymmetry via out-of-equilibrium decays, $N
\rightarrow \ell H$ and $N \rightarrow \overline{\ell} \,
\overline{H}$, where $H$ is the SU(2) Higgs doublet and $\ell$ is the
lepton doublet. The quantum interference between the tree-level and
one-loop contributions to the decay can lead to a lepton number
asymmetry.  

The predictions for the baryon number asymmetry through standard
leptogenesis depends on the coupling constants in the seesaw
Lagrangian. Thus, by demanding that sufficient baryon number asymmetry
to be generated, constraints~\cite{DiBari:2012fz} on neutrino
parameters can be obtained. Specifically, the mass of the lightest
right-handed neutrino is constrained to be $M_{1} >~ 3 \times
10^{9}$ GeV. One also obtains an upper bound on the light effective
neutrino mass, $m_{1} < 0.12$ eV, which is incompatible with the
quasi-degenerate spectrum.  

If the right-handed neutrinos are produced thermally, the lower bound
on the lightest right-handed neutrino mass is then translated into a
lower bound on the reheating temperature, $M_{\rm {RH}} > M_{1} >
O(10^{9})$ GeV. Such a high reheating temperature is problematic for
many extensions of the Standard Model. For instance in many variants
of supersymmetry, constraints from WMAP (for stable gravitino) and BBN
(for unstable gravitino) typically require the reheating temperature
to be several orders of magnitude lower than $10^{9}$ GeV,
incompatible with the condition for successful standard leptogenesis
in supersymmetric models.  

\subsubsection{Alternative Realizations} 

To evade the gravitino over production problem, several scenarios have
been proposed in which the conflicts between leptogenesis and
gravitino over-production problem are overcome in different ways:
\begin{itemize}
\item resonant enhancement in the self-energy diagrams due to near
  degenerate right-handed neutrino masses: in resonant
  leptogenesis~\cite{Pilaftsis:2003gt}, it has been shown that
  sufficient asymmetry can be generated even with TeV scale
  right-handed neutrino masses, leading to the possibility of testing
  this scenario at the collider experiments.
\item relaxing the relation between the lepton number asymmetry and
  the right-handed neutrino mass: one example is the soft
  leptogenesis~\cite{Grossman:2003jv}, where the asymmetry arises in
  mixing, instead of decay. In this case, the source of CP violation
  is the complex phases in the soft SUSY parameters.
\item relaxing the relation between the reheating temperature and the
  right-handed neutrino mass: one realization of non-thermal
  leptogenesis is the production of the right-handed neutrinos by
  inflaton decay~\cite{Asaka:1999yd}.
\end{itemize}

\subsubsection{Dirac Leptogenesis} 

It was pointed out ~\cite{Dick:1999je} that leptogenesis can be
implemented even in the case when neutrinos are Dirac fermions which
acquire small masses through highly suppressed Yukawa couplings
without violating lepton number. The realization of this depends
critically on the following three characteristics of the sphaleron
effects: ({\it i}) only the left-handed particles couple to the
sphalerons; ({\it ii}) the sphalerons change (B+L) but not (B-L);
({\it iii}) the sphaleron effects are in equilibrium for $T >~ 
T_{EW}$. For the neutrinos, given that the neutrino Dirac mass is very
tiny ($m_{D} < 10$ keV), the left-right equilibration can occur at a
much longer time scale compared to the electroweak epoch when the
sphaleron washout is in effect. Suppose that some processes initially
produce a negative lepton number ($\Delta L_{L}$), which is stored in
the left-handed neutrinos, and a positive lepton number ($\Delta
L_{R}$), which is stored in the right-handed neutrinos. Because
sphalerons only couple to the left-handed particles, part of the
negative lepton number stored in left-handed neutrinos get converted
into a positive baryon number by the electroweak anomaly. This
negative lepton number $\Delta L_{L}$ with reduced magnitude
eventually equilibrates with the positive lepton number, $\Delta
L_{R}$ when the temperature of the Universe drops to $T \ll
T_{EW}$. Because the equilibrating processes conserve both the baryon
number $B$ and the lepton number $L$ separately, they result in a
Universe with a total positive baryon number and a total positive
lepton number. And hence a net baryon number can be generated even
with $B=L=0$ initially.  

\subsubsection{Possible connections to CP violation in neutrino oscillation.} 

In the seesaw Lagrangian at high scale in the presence of three
right-handed neutrinos, there are in total 6 mixing angles and 6
physical CP phases. On the other hand, the effective Lagrangian at low
energy after integrating out the right-handed neutrinos, only three
mixing angles and three physical CP phases remain. Given the presence
of extra mixing angles and phases at high energy, it is generally
impossible to connect leptogenesis (within the standard leptogenesis
framework) and low energy CP violation processes in a model
independent way. Nevertheless, this statement is weakened when the
flavor effects, which are relevant if leptogenesis takes place at $T <
10^{12}$ GeV, are taken into account. On the other hand, within
certain predictive models for neutrino masses, strong connections can
be established even in the absence of flavor effects.  

Generally, two classes of models have been shown to exhibit possible
connection between leptogenesis and CP violation in neutrino
oscillation. These include:
\begin{itemize}
\item models with rank-2 mass matrix: It has been shown that in a
  model with only two right-handed neutrinos with a rank-2 neutrino
  mass matrix, the sign of the baryon number asymmetry is related to
  the sign of CP violation in neutrino
  oscillation~\cite{Frampton:2002qc}.
\item models where CP violation comes from a single source: These
  include models with spontaneous CP violation, for example in minimal
  left-right symmetry model, there is only one physical CP phase in
  the lepton sector~\cite{Chen:2004ww}. All leptonic CP violating
  processes (leptogenesis, neutrino oscillation, etc) are determined
  solely by this phase. Another example is a model with finite group
  family symmetry $T^{\prime}$ with complex Clebsch-Gordan
  coefficients.  CP violation in this model is due entirely to the
  complex CG coefficients. As the only non-vanishing leptonic phases
  are the low energy ones due to the symmetry of the model, there
  exists a strong connection between leptogenesis and low energy CP
  violating processes in this model~\cite{Chen:2011tj}.
\end{itemize}

\subsubsection{Affleck-Dine Baryogenesis} 

In supersymmetric extensions of the Standard Model, there exist field
configurations--condensates-- of squark and slepton fields with very
large expectation values and relatively low energy density compared
with the thermal energy. In supersymmetric theories, at the exit from
inflation, the observable universe would typically be in one of these
configurations. Affleck and Dine showed that $CP$ and baryon number
violation at high energy would lead to net baryon production from the
coherent evolution and subsequent decay of the condensates
\cite{Affleck:1984fy}. The Affleck-Dine scenario is consistent with a
low reheat scale, and in some variants, can provide an explanation for
dark matter as well. Decay of the condensates into baryons and WIMPS
can provide an explanation for the similar cosmological densities of
dark matter and baryons\cite{Thomas:1995ze,Enqvist:1998en}.  Another
interesting possibility is that the condensates will fragments into
stable lumps of matter with macroscopic amounts of baryon number and
lower energy/baryon number than ordinary matter. For some parameters,
these lumps, called Q-balls, are a viable dark matter candidate with
unusual phenomenology
\cite{Kusenko:1997si,Kusenko:1997vp,Kusenko:2005du}.

\subsubsection{Electroweak Baryogenesis} 

 A strongly first order electroweak phase transition can occur in some
 extensions of the standard model, and provide the departure from
 thermal equilibrium necessary for baryogenesis.  Such a phase
 transition proceeds via nucleation of bubbles of broken phase which
 expand to fill the entire universe.  Inside the bubbles the Higgs
 expectation value is large, sphaleron transitions are suppressed, and
 baryon number is conserved, while the symmetric phase with no Higgs
 expectation value and unsuppressed sphalerons exists outside the
 bubbles. CP violating scattering of particles with the expanding
 bubble walls can lead to an CP asymmetric particle content in the
 symmetric phase, which will bias the sphalerons towards producing a
 net baryon number. This baryon number will then pass into the bubbles
 and, provided the sphalerons inside the bubble are suppressed by a
 large enough Higgs expectation value, survive until the present. For
 strongly first order phase transition, the effective potential at the
 critical temperature for the Higgs field must possess degenerate
 minima with a barrier between them. This barrier requires new bosons
 which are coupled to the Higgs field. Some well explored contenders
 for electroweak baryogenesis models are the MSSM \cite{Cohen:1992zx},
 and generic two Higgs doublet models\cite{Turok:1990zg}. The MSSM is
 still a viable baryogenesis model provided the scalar partner of the
 right handed top quark is lighter than the top
 quark\cite{Carena:1996wj}, and nonminimal supersymmetric theories are
 much less constrained\cite{Pietroni:1992in}. In general two Higgs
 doublet models have a large parameter
 space\cite{Cline:1996mga,Fromme:2006cm}, a significant portion of
 which remains viable for baryogenesis after the recent 126 GeV Higgs
 boson discovery\cite{Dorsch:2013wja}. Both the MSSM and two Higgs
 doublet models possess potential additional sources of CP violation
 and new contributions to Electric Dipole Moments (EDMs), which
 constrain the magnitude of the new CP violating phases.  The theory
 connecting the new sources of CP violation with the total baryon
 number produced is very complicated and still possesses considerable
 sources of uncertainty, and is still being actively developed.  See
 ref. \cite{Morrissey:2012db} for a recent review. Eventually reliable
 theoretical computations of the baryon asymmetry will allow for
 predictions for EDMs and for new particle properties in electroweak
 baryogenesis models, but we are not quite there yet.
 
\subsubsection  {Other Baryogenesis mechanisms} 

Sakharov's original model, and subsequent baryogenesis models based on
Grand Unified Theories, relied on the out of equilibrium decays of
very heavy particles of mass of order $10^{15}$ GeV. Such theories are
mostly now inconsistent within the modern theory of inflation, as it
is difficult to obtain a high enough reheat temperature to produce
such particles.
  
  A variety of other mechanisms for baryogenesis have been suggested,
  at energy scales ranging from just above the nucleosynthesis
  temperature of an MeV \cite{Dimopoulos:1987rk} to very high
  temperatures or high scale out of equilibrium processes occurring at
  the end of inflation. Some theories do not require baryon number
  violation at all, as the dark matter can carry an equal and opposite
  baryon number \cite{Dodelson:1989cq}.
  
\subsection  {Experimental Signatures of Baryogenesis}
  
The origin of the matter---anti-matter asymmetry in the universe is
one of the most profound scientific questions of our time.  A wide
variety of experiments at all three frontiers can provide
illumination.

\subsubsection{Cosmic Frontier Experiments and Baryogenesis:} 
\begin{itemize}
\item Constraining the scale of inflation from the impact of tensor
  fluctuations on the CMB is important. \item Gravity wave experiments
  could provide evidence for a first order phase transition.
\item Certain dark matter models are connected with baryogenesis and
  can provide unusual signatures.
\end{itemize}
\subsubsection {Intensity Frontier experiments and Baryogenesis:} 
 \begin{itemize}
\item
CP violation in the neutrino sector would provide support for the
leptogenesis scenario, and some models make for specific
predictions. \item Evidence for or against neutrino Majorana masses
would also impact leptogenesis theory.
\item Electroweak baryogenesis scenarios provide strong motivation to
  search for EDMs, exotic CPV in meson physics, and rare decays.
\item Proton d`qecay provides an important constraint on Grand Unified Model Building and on new sources of Baryon number violation.
\end{itemize}
\subsubsection{\bf Energy Frontier experiments and Baryogenesis:} 
 \begin{itemize}
 \item Collider tests of extended Higgs sectors and searches for new
   light scalars are important.
 \item Electroweak baryogenesis models typically have sizeable
   modifications of the triple Higgs self coupling
   \cite{Kanemura:2004ch,Noble:2007kk}
 \end{itemize}

%% file: CF6/holometer.tex
\subsection{Quantum Geometry and The Holographic Universe}

New quantum degrees of freedom of space-time, originating at the
Planck scale, could create a coherent indeterminacy and noise in the
transverse position of massive bodies on macroscopic scales.
 
Quantum effects of space-time are predicted to originate at the Planck
scale, $ct_P\equiv \sqrt{\hbar G/c^3}= 1.616\times 10^{-35}$m.  In
standard quantum field theory, their effects are strongly suppressed
at experimentally accessible energies, so space-time is predicted to
behave almost classically, for practical purposes, in particle
experiments.  However, new quantum effects of geometry originating at
the Planck scale--- from geometrical degrees of freedom not included
in standard field theory--- may have effects on macroscopic scales
that could be measured by laser interferometers.

The possibility of new quantum-geometrical degrees of freedom is
suggested from several theoretical directions. Quantum physics is
experimentally proven to violate the principle of locality on which
classical space-time is based. Gravitational theory suggests that
quantum states of space-time systems do not respect locality of the
kind assumed by quantum field theory, and suggests that space-time and
gravity are approximate statistical behaviors of a quantum system with
a holographic information content, far less than that predicted by
quantum field theory.\cite{Jacobson:1995ab,Verlinde:2010hp}

Quantum geometry could arise in Planck scale physics, but still
produce a detectable displacement in a macroscopic
experiment.\cite{Hogan:2012ib} A typical uncertainty in wave
mechanics, if information about transverse position is transmitted
nonlocally with a bandwidth limit, is the scale familiar from
diffraction-limited imaging: the geometric mean of inverse bandwidth
and apparatus size. For separations on a laboratory scale, a Planck
scale frequency limit leads to a transverse uncertainty in position on
the order of attometers.  Displacements of massive bodies of this
order are detectable using laser interferometry.

No fundamental theory of quantum geometry exists, but a consistent
effective theory, based on general properties of quantum mechanics and
covariance, can be used to precisely predict a phenomenology on
macroscopic scales.  In particular, the theory precisely relates the
number of geometrical position eigenstates to the amplitude of
indeterminacy in transverse position at separation $L$, so it can be
related to the holographic density of states predicted from
gravitational theory.  This hypothesis leads to an exact prediction
for the variance in transverse position with no free
parameters,\cite{Hogan:2012ne}
\begin{equation}
\langle x_\perp^2 \rangle= L ct_P/ \sqrt{4\pi} .
\end{equation}
Planckian indeterminacy leads to a new form of noise in position with
this displacement, on a timescale $L/c$.  This form of indeterminacy
would have escaped detection to date, and indeed is overwhelmed by
standard quantum indeterminacy on the mass scale of elementary
particles.  However, it is detectable as a new source of
quantum-geometrical noise in an interferometer that coherently
measures the positions of massive bodies in two directions over a
macroscopic volume.\cite{Hogan:2010zs,Hogan:2012ib}

\subsection{The Fermilab Holometer}

An experiment is under development at Fermilab designed to detect or
rule out a transverse position noise with Planck spectral density,
using correlated signals from an adjacent pair of Michelson
interferometers. A detection would open an experimental window on
quantum space-time.

The Fermilab Holometer is an experiment (E-990) designed to
detect or rule out quantum-geometrical noise with these
properties.\cite{holowebsite} Much of the technology has been
developed by LIGO and other projects to measure displacements due to
gravitational radiation.  The quantum-geometrical measurement however
calls for application of the technology in a new experimental design.
Measurements can be made at relatively high (MHz) frequencies, where
environmental and gravitational noise sources are smaller, both
shrinking and simplifying the layout. The experiment is designed to
measure the specific and peculiarly quantum-mechanical signatures of
the effect, such as nonlocal coherence and transverse nature of the
indeterminacy, the frequency cross spectrum, and time-domain cross
correlation function. It is anticipated that the experiment will be
complete, and either detect or rule out this form of Planckian noise,
within about two years.

If the noise is found not to exist, only a modest followup effort may
be motivated to pursue the limits somewhat past the Planck scale for a
conclusive result. If it is found, a significantly expanded
experimental program can be pursued to obtain high precision results
and map out the spatiotemporal properties of quantum geometry.

\subsection{Torsion Balance Experiments}

Another example of table top experiments with sensitivity to quantum gravitational effects are torsion balance experiments \cite{Adelberger:2013jwa}, which are sensitive to preferred frame effects, noncommutative geometry, new dimensions at short distances, and equivalence principle violation.

%% file: CF6/TQ.tex
\section{Tough Questions}

\noindent {\it CF34. What are the roles of cosmic-ray, gamma-ray, and
  neutrino experiments for particle physics? What future experiments
  are needed in these areas and why? Are there areas in which these
  can have a unique impact?}

In this document we have discussed a broad range of fundamental physics that can be gleaned from cosmic rays, gamma rays, and neutrinos (dark matter, axions, primordial black holes, Q-balls, Lorentz invariance violation, intergalactic magnetic fields, particle interactions at high energies, neutrino mass hierarchy, etc.) that can not be reached through other methods.   

{\bf \underline{CTA}} will usher in the era of precision VHE astrophysics and in conjunction with current instruments (Fermi and HAWC) will provide a view of the high-energy universe that will lead to an understanding of the astrophysical processes at work in these extreme objects and enable us to probe the laws of physics at energies, couplings, and mass scales that are beyond the reach of traditional high-energy physics experiments.  Most importantly perhaps is the indirect detection of dark matter (see CF-2), where gamma-ray experiments have sensitivity to regions of parameter space not accessible to other techniques (accelerator and direct search techniques).   The discovery of a primordial black hole or a Q-ball would would provide a wealth of data on the early universe and particle physics at energies not attainable in accelerators.  Measurement of a vacuum dispersion relation for light would provide unique insight into the merging of quantum mechanics and general relativity.  A measurement of the intergalactic magnetic fields would give insight into primordial magneto-genesis and the early universe processes, phase transitions or inflation dynamics that may have given rise to such a field.

{\bf \underline{PINGU}} will use the atmospheric neutrinos to measure the neutrino mass hierarchy.  By using the atmospheric neutrinos generated by cosmic-ray interactions in the atmosphere a large range of $L/E$ is available, enabling sensitive searches for matter effects on neutrino oscillations.  At a relatively low cost, this can be used to make a definitive measurement of the neutrino mass hierarchy.

{\bf \underline{JEM-EUSO and a large aperture ground array}}  Cosmic-ray experiments provide a window into particle interactions at the highest energies.  While only large cross-section physics is accessible, given the low flux, well measured data with well-understood systematics, may offer channels to new physics.  With much improved control over systematics and detector resolution and by combining various detectors (hybrid approach) current data has yielded interesting hints that we don't fully understand how to predict hadronic interactions at these energies.  The measurement of total cross-section by HiRes and Auger at energies well beyond the LHC, while still crude, can be refined and be an important constraint on hadronic interaction models.  A future large area surface detector will extend this measurement to higher energies, over an order of magnitude greater than achievable at the LHC.

From an astrophysics point of view, the next big question is the anisotropy of UHECR, i.e. the search for astrophysical sources. This is beginning to show interesting hints of association with nearby Large Scale Structure, but to really pin this down will require another order of magnitude or more of collecting area.   JEM-EUSO, can achieve the required exposure.  We know that UHECR originate within $\sim$100 Mpc of us.  But until we can clearly state that the highest energy cosmic-ray flux anisotropy is understood from the point of view of astrophysical sources, the potential for new physics remains as strong as ever.

\noindent {\it CF35. What will it take to identify the mechanism for
  baryogenesis or leptogenesis? Are there scenarios that could
  conceivably be considered to be established by experimental data in
  the next 20 years?}

There are 3 well motivated scenarios for baryogenesis. Of these, electroweak baryogenesis is the one which is most likely to be definitively established or excluded within 20 years.  In this scenario the baryon asymmetry of the universe can be related to the properties of new particles which couple to the Higgs boson, to properties of the Higgs boson, and to electric dipole moments (EDMs) of the neutron, the electron, and atoms. In particular difficult but doable theoretical calculations can relate the size and sign of a new CP violating phase to the baryon asymmetry and to EDMs.

It will be more difficult to establish the leptogenesis scenario, especially in the versions where it proceeds via the decay of very heavy ($>10^9$ GeV) neutrinos. Some generic indications of this scenario are Majorana neutrinos, a light neutrino below 0.1 eV, and CP violation in neutrino oscillations, but verification of these features does not prove the theory is right, and the theory does not make a specific prediction for the phase which is observable in oscillations. However the heavy neutrino version requires a high inflation scale and high reheat temperature after inflation, which could be inconsistent with some kinds of new physics such as neutron-anti-neutron oscillations or supersymmetry with a gravitino in the mass range between a keV and $10^4$ GeV, and so, depending on what other new physics is discovered, it could be excluded. Other leptogenesis scenarios involve new neutral leptons at or below the weak scale, or Dirac neutrinos, and have a restricted enough parameter space to be excluded or confirmed.

Affleck-Dine baryogenesis requires supersymmetry but does not necessarily require a specific SUSY spectrum, so definitively establishing or excluding this scenario is difficult. It is possible that this scenario can lead to the formation of stable Q-balls, an interesting dark matter candidate whose discovery would be strong evidence for the Affleck Dine scenario. Evidence of a high inflation scale from CMB B-modes would likely exclude this scenario, as fluctuations in the condensate would lead to isocurvature perturbations with an amplitude that has been ruled out by CMB measurements.

\noindent {\it CF36. What are the leading prospects for detecting GZK
  neutrinos?  What experimental program is required to do this in the
  next 5 years, 10 years, 20 years, and how important is this?}

The prospects for detecting GZK neutrinos in the next decade are quite good.  We know that there is a bottom to the flux of these neutrinos.   Current efforts have ruled out the more optimistic scenarios and are now reaching into realistic parameter space.  The next generation of instruments can improve upon current sensitivity by an order of magnitude giving sensitivity that is close to the lowest possible fluxes (an all iron composition).  At the same time, the fraction of iron in the highest energy cosmic rays is lower than previously thought, significantly increasing the likelihood that the next generation of experiments will detect the GZK neutrinos.  

How important is this?  The GZK neutrinos provide a method to measure the neutrino cross section at a center-of-mass energy of 100 TeV!  Observation of these interactions can provide information on beyond standard model physics that is many orders of magnitude beyond what is achievable in accelerator-based neutrino experiments.  These observations will also enable us to understand cosmic accelerators.